\documentclass[preprint,pre,showpacs]{revtex4}
\usepackage{amscd}
\usepackage{amsfonts}
\usepackage{amsmath}
\usepackage{amssymb}
\usepackage{undertilde}
\usepackage{color}
\usepackage{graphicx}
\usepackage{graphics}
\usepackage{wrapfig}
\usepackage[T1]{fontenc}
\usepackage[latin1]{inputenc}
\usepackage{gensymb}
\def\>{\rangle}
\def\<{\langle}
\def\n{\nonumber}

\def\sc{\scriptsize}
\newcommand{\id}{\openone}
\begin{document}
\title{Non-negative Wigner-like distributions and R\'{e}nyi-Wigner entropies of arbitrary non-Gaussian quantum states:
The thermal state of the one-dimensional box problem}
\author{Ilki Kim}
\email{hannibal.ikim@gmail.com} \affiliation{Joint School of
Nanoscience and Nanoengineering, North Carolina A$\&$T State
University, Greensboro, NC 27411}
\date{\today}
\begin{abstract}
In this work, we consider the phase-space picture of quantum
mechanics. We then introduce non-negative Wigner-like (operational)
distributions $\widetilde{\mathcal W}_{\rho;\alpha}(x,p)$
corresponding to the density operator $\hat{\rho}$ and being
proportional to $\{W_{\rho^{\scriptscriptstyle \alpha/2}}(x,p)\}^2$,
where $W_{\rho}(x,p)$ denotes the usual Wigner function. In doing
so, we utilize the formal symmetry between the purity measure
$\mbox{Tr}(\hat{\rho}^2)$ and its Wigner representation $(2\pi\hbar)
\int dx dp\, \{W_{\rho}(x,p)\}^2$ and then consider, as a
generalization, such symmetry between the fractional moment
$\mbox{Tr}(\hat{\rho}^{\alpha})$ and its Wigner representation
$(2\pi\hbar) \int dx dp\, \{W_{\rho^{\scriptscriptstyle
\alpha/2}}(x,p)\}^2$. Next, we create a framework that enables
explicit evaluation of the R\'{e}nyi-Wigner entropies for the
classical-like distributions $\widetilde{\mathcal
W}_{\rho;\alpha}(x,p)$. Consequently, a better understanding of some
non-Gaussian features of a given state $\hat{\rho}$ will be given,
by comparison with the Gaussian state $\hat{\rho}_{\mbox{\tiny G}}$
defined in terms of its Wigner function $W_{\rho_{\mbox{\tiny
G}}}(x,p)$ and essentially determined by its purity measure
$\mbox{Tr}\{(\hat{\rho}_{\mbox{\tiny G}})^2\}$ alone. To illustrate
the validity of our framework, we evaluate the distributions
$\widetilde{\mathcal W}_{\beta;\alpha}(x,p)$ corresponding to the
(non-Gaussian) thermal state $\hat{\rho}_{\scriptscriptstyle \beta}$
of a single particle confined by a one-dimensional infinite
potential well with either the Dirichlet or Neumann boundary
condition and then analyze the resulting R\'{e}nyi entropies. Our
phase-space approach will also contribute to a deeper understanding
of non-Gaussian states and their properties either in the
semiclassical limit ($\hbar \to 0$) or in the high-temperature limit
($\beta \to 0$), as well as enabling us to systematically discuss
the quantal-classical Second Law of Thermodynamics on the single
footing.
\end{abstract}
\maketitle
%
\section{Introduction}\label{sec:introduction}
%
The R{\'{e}nyi-$\alpha$ entropy of a probability distribution $P =
\{p_j\}$ was originally introduced in classical information theory,
explicitly given by $S_{\alpha}(P) = (1-\alpha)^{-1} \ln\{\sum_j
(p_j)^{\alpha}\}$ with order $\alpha
> 0$ \cite{REN61,REN65}, as a generalization of the Shannon measure of information (SMI) given by $S_1(P)
= -\sum_j p_j \ln p_j$ (for a deeper conceptual discussion of SMI,
see, e.g., \cite{BEN17}). Then, its quantum analog, given by
$S_{\alpha}(\hat{\rho}) = (1 - \alpha)^{-1}
\ln\{\mbox{Tr}(\hat{\rho}^{\alpha})\}$ of a density operator
$\hat{\rho}$, has been studied in various contexts of quantum
information theory and quantum thermodynamics
\cite{WEH78,LIN13,MIS15,HAM16,ABE16,KIM18,DON19}, which is
accordingly a generalization of the von Neumann entropy
$S_1(\hat{\rho}) = -\mbox{Tr}(\hat{\rho}\,\ln\hat{\rho})$; for
instance, for a generalized formulation of quantum thermodynamics,
which is built upon the maximum entropy principle applied to
$S_{\alpha}(\hat{\rho})$ \cite{MIS15}, as well as in the discussion
of its time derivative under the Lindblad dynamics, the result of
which may be useful for exploring the dynamics of quantum
entanglement in the Markovian regime \cite{ABE16}. Obviously, the
case of $\alpha = 2$ gives the well-known quantity $S_2(\hat{\rho})
= -\ln\{\mbox{Tr}(\hat{\rho}^2)\}$.

The quantum-mechanical expectation values of observables $\hat{A}$
can be calculated independently of the pictures in consideration
(i.e., either the operator picture in the Hilbert space or the
$c$-number picture in the classical phase space), and are required
to obey, e.g., the Ehrenfest theorem (as a quantum-classical
channel) stating that the classical laws of motion also hold true
formally for the quantal expectation values \cite{GRI05}. On the
other hand, the density operator $\hat{\rho}$ itself of the
Hilbert-space picture can possess genuine quantum features (or
non-classicalities) such as coherence and entanglement, which have
been attracting considerable interest, as the need for a better
theoretical understanding of them increases in response to the
experimental manipulation of them in small quantum systems
\cite{CHU00,MAH04}. However, when it comes to a systematic study of
the canonical quantum-classical transition in the limit of $\hbar
\to 0$, it would be more favorable in some contexts to take into
consideration the quasi-probability distribution of the $c$-number
picture, such as the Wigner function
\cite{WIG32,HIL84,LEE95,SCH01,CUR05}
\begin{subequations}
\begin{eqnarray}\label{eq:bopp_0-1}
    W_{\rho}(x,p) &=& \frac{1}{\pi\hbar} \int_{-\infty}^{\infty} d{\xi}\,
    \left\<x + \xi\right|\hat{\rho}\left|x - \xi\right\>\, \exp\left(-\frac{2i}{\hbar}
    p\,\xi\right)\\
    \hat{\rho} &=& 2 \int d\xi \int dx dp\, |x+\xi\>\, W_{\rho}(x,p)
    \exp\left(\frac{2i}{\hbar}
    p\,\xi\right) \<x-\xi|\,.\label{eq:rho_from wigner_1}
\end{eqnarray}
\end{subequations}
As such, the Wigner function $W_{\rho}(x,p)$ may be regarded as the
direct counterpart to the classical probability distribution
$P(x,p)$, and therefore the quantum-classical channels between the
two $c$-number distributions can also be explored. However, the
Wigner function of a generic quantum state may possess negative
values, as is well-known.

Therefore, it will be interesting information-theoretically and
thermodynamically to discuss the quantal-classical Second Law on the
single footing, i.e., in terms of the R\'{e}nyi-$\alpha$ entropy
$S_{\alpha}(W_{\rho})$ for a given distribution $W_{\rho}(x,p)$,
albeit with its negative values, in addition to the First Law,
expressed in terms of the internal energy $\<\hat{A}\>$ with
$\hat{A} \to \hat{H}$, where the expectation value is given by
\begin{equation}\label{eq:expectation1}
    \<\hat{A}\> = \mbox{Tr}(\hat{\rho} \hat{A}) = \int dx dp\; W_{\rho}(x,p)\; W_{A}(x,p)
\end{equation}
with the Weyl-Wigner representation
\begin{equation}\label{eq:bopp_0-2}
    W_{\scriptscriptstyle A}(x,p)\, =\, \int_{-\infty}^{\infty} d{\xi}\,
    \left\<x + \frac{\xi}{2}\right|\hat{A} \left|x - \frac{\xi}{2}\right\>\, \exp\left(-\frac{i}{\hbar}
    p\,\xi\right)\,.
\end{equation}
Thus far, such a unified approach of the R\'{e}nyi-$\alpha$
entropies has not extensively been discussed, except for either the
case of $\alpha = 2$, for which the purity measure
$\mbox{Tr}(\hat{\rho}^2) = (2\pi\hbar) \int dx
dp\,\{W_{\rho}(x,p)\}^2$ given in compact form, or the case of a
Gaussian state $\hat{\rho}_{\mbox{\tiny G}}$ (due to its
mathematically simple structure) defined as a quantum state, the
Wigner function $W_{\rho_{\mbox{\tiny G}}}(x,p)$ of which is
Gaussian, such as the canonical thermal equilibrium state
$\hat{\rho}_{\beta}$ (with $\beta = 1/k_{{\mbox{\sc B}}} T$) of an
$N$-oscillator system (corresponding to its ground state at $T =
0$), the coherent state, and the squeezed state, etc.
\cite{BRA05,FER05,WOL06,WED12,OLI12,ADE14}. Further, the entropy
$S_2(W_{\rho_{\mbox{\tiny G}}})$ of a Gaussian state has been shown
to coincide with the so-called Wigner entropy $S_{\mbox{\tiny
W}}(W_{\rho_{\mbox{\tiny G}}})$ up to a constant, where
$S_{\mbox{\tiny W}}(W_{\rho_{\mbox{\tiny G}}}) := -\int dx dp\,
W_{{\rho}_{\mbox{\tiny
G}}}(x,p)\,\ln\{(2\pi\hbar)\,W_{{\rho}_{\mbox{\tiny G}}}(x,p)\}$ is
well-defined due to the non-negativity of $W_{{\rho}_{\mbox{\tiny
G}}}(x,p)$ over the entire phase space \cite{ADE12,PAT17}.

It has also been known \cite{GEN10,ADE14} that while Gaussian states
are crucial resources for quantum information processing (QIP) with
continuous variables, non-Gaussian states also are either required
or desirable (in terms of efficiency) for some tasks relevant to QIP
such as entanglement distillation \cite{EIS02,GIE02,FIU02}, cluster
quantum computation \cite{RAL03,LUN08}, and teleportation
\cite{DEL07,WAN15}, etc. In doing so, the so-called non-Gaussianity
(nG) was used as a critical resource of the information processing,
and some measures of nG have then been proposed for quantification
of the non-Gaussian character of a given state
$\hat{\rho}_{\mbox{\tiny nG}}$, such as the measure defined in terms
of the Hilbert-Schmidt distance between $\hat{\rho}_{\mbox{\tiny
nG}}$ and its reference Gaussian state $\hat{\sigma}_{\mbox{\tiny
G}}$ with the same first and second moments of the canonical
quadrature operators as $\hat{\rho}_{\mbox{\tiny nG}}$ \cite{GEN07},
the quantum relative entropy (QRE) between $\hat{\rho}_{\mbox{\tiny
nG}}$ and $\hat{\sigma}_{\mbox{\tiny G}}$ \cite{GEN08,MAR13}, and
the one defined in terms of the difference between the Wehrl
entropies of $\hat{\rho}_{\mbox{\tiny nG}}$ and
$\hat{\sigma}_{\mbox{\tiny G}}$ \cite{IVA12}, as well as the one
defined in terms of the Bures distance between
$\hat{\rho}_{\mbox{\tiny nG}}$ and $\hat{\sigma}_{\mbox{\tiny G}}$
\cite{GHI13}.

However, each measure of nG has its drawback; e.g., a drawback of
the QRE, used most widely for a general (non-Gaussian) state
$\hat{\rho}$, is that its actual evaluation requires the full
information about $\hat{\rho}$ such that it is often not feasible to
calculate if only partial information is available \cite{GEN10}.
Therefore, it will also be desirable to discuss the
R{\'{e}nyi-$\alpha$ entropies $S_{\alpha}(\hat{\rho})$ in this
context, which will provide the higher-moment information of
$\hat{\rho}$ beyond $\mbox{Tr}(\hat{\rho}^2)$, thus enabling us to
approximately determine the state as accurately as possible, while
Gaussian states (``tame'' in their behaviors) are essentially
determined by $S_2(\hat{\rho})$ alone \cite{OLI12,ADE14,KIM18}. As
such, the entropies $S_{\alpha}(\hat{\rho})$ of a non-Gaussian state
may contain the nG information in the operational sense. Further,
the non-Gaussian features of $\hat{\rho}$, as the deviation from
those of the Gaussian state $\hat{\rho}_{\mbox{\tiny G}}$ being
defined in terms of its Wigner function $W_{{\rho}_{\mbox{\tiny
G}}}(x,p)$, give impetus to a discussion of the entropies
$S_{\alpha}(\hat{\rho})$ (with $\alpha \in {\mathbb R}^+$) in the
phase-space picture, which will, in turn, enable us to explore more
rigorously their quantum-classical transitions in the context of
information theory and thermodynamics.

Exact evaluations of R\'{e}nyi-$\alpha$ entropies
$S_{\alpha}(W_{\rho})$ for given distributions $W_{\rho}(x,p)$ of
non-Gaussian states have been studied in \cite{KIM18}. The resulting
framework has successfully provided a general expression for
calculating the entropies $S_{l}(W_{\rho})$ for integer orders $l$.
Within this framework, the entropies
$S_{\alpha}(W_{\rho_{\mbox{\tiny G}}})$ of Gaussian states
$W_{\rho_{\mbox{\tiny G}}}(x,p)$ for real values of $\alpha
> 0$ also have been rediscovered in closed form, but with the help of an additional recurrence relation
between two consecutive entropies $S_l$ and $S_{l+1}$ followed by
the analytic continuation of $l \to \alpha$. However, it still
remains an open question to directly evaluate the entropies
$S_{\alpha}(W_{\rho_{\mbox{\tiny nG}}})$ of non-Gaussian states
$W_{\rho_{\mbox{\tiny nG}}}(x,p)$ for real values $\alpha$ for which
finding such a recurrence relation would be a formidable task.

In this work, we intend to create another framework, as a
generalization of the preceding one, in which a group of Wigner-like
(operational) distributions denoted by $\widetilde{\mathcal
W}_{\rho;\alpha}(x,p)$ (with order $\alpha$ including the case of
$\alpha = 2$, obviously) will be introduced, corresponding to the
same density operator $\hat{\rho}$, and then the R\'{e}nyi entropies
$S_{\alpha}(\widetilde{\mathcal W}_{\rho;\alpha})$, being tantamount
to $S_{\alpha}(\hat{\rho})$, can be evaluated in the phase-space
picture for arbitrary non-Gaussian states, actually with no need for
the aforesaid recurrence relation and analytic continuation [cf.
Eqs. (\ref{eq:tilde_wigner1}) and (\ref{eq:wigner-entropy_2})].
Remarkably, the distributions $\widetilde{\mathcal
W}_{\rho;\alpha}(x,p)$ will be shown to be non-negative over the
entire phase space (like the classical probability distribution) and
well-defined in the genuine quantum regime all the way to the
semiclassical limit. Besides, because of the equivalence between
$S_{\mbox{\tiny W}}(W_{\rho_{\mbox{\tiny G}}})$ and
$S_2(W_{\rho_{\mbox{\tiny G}}})$ (up to a constant) for
(non-negative) Gaussian states and also the non-negative feature of
$\widetilde{\mathcal W}_{\rho;\alpha}(x,p)$ for non-Gaussian states,
the resulting entropies $S_{\alpha}(\widetilde{\mathcal
W}_{\rho;\alpha})$, called the R\'{e}nyi-Wigner entropies, may also
be regarded as a generalization of the Wigner entropy
$S_{\mbox{\tiny W}}(W_{\rho_{\mbox{\tiny G}}})$. Subsequently, we
will consider a specific non-Gaussian state that will be applied for
our framework of its R\'{e}nyi-Wigner entropies; this is the thermal
state $\hat{\rho}_{\beta}$ of a single particle confined by a
one-dimensional infinite potential well with either Dirichlet or
Neumann boundary condition. Its Wigner function (with its negative
values) will be shown to tend asymptotically to a Gaussian shape in
the limit of $\hbar \to 0$ only.

The general layout of this paper is as follows: In Sec.
\ref{sec:wigner entropy} we introduce a group of Wigner-like
distributions as variants of the Wigner function and then provide a
generic framework for the R\'{e}nyi-Wigner entropies of arbitrary
quantum states in the classical phase space. In Sec. \ref{sec:sec3}
we explicitly evaluate the Wigner function, and its variants, of the
thermal state of the one-dimensional box problem and then discuss
the relevant issues of quantum-classical transition. In Sec.
\ref{sec:sec4} we apply our framework for this thermal state, and
discuss some subjects relevant to the resulting R\'{e}nyi-Wigner
entropies. Finally, we provide concluding remarks in Sec.
\ref{sec:conclusion}.

\section{Non-negative Wigner-like distributions and entropies}\label{sec:wigner entropy}
%
We first observe that the purity measure, being the first moment of
probability $\<p\> = \sum_n\,(p_n)^2$ with the eigenvalues $p_n$'s
of $\hat{\rho}$, may be rewritten as
\begin{equation}\label{eq:purity_measure1}
    \mbox{Tr}(\hat{\rho}^{\scriptscriptstyle 2}) = (2\pi\hbar) \left\<W_{\rho}\right\>_{\scriptscriptstyle W_{\scriptscriptstyle
    \rho}} = \int dx dp\, {\mathcal W}_{\rho^2}(x,p)\,,
\end{equation}
in which the expectation value
$\left\<W_{\rho}\right\>_{\scriptscriptstyle W_{\scriptscriptstyle
\rho}} := \int dx dp\, W_{\rho}(x,p)\, W_{\rho}(x,p)$, and the
symbol ${\mathcal W}_{\rho^{\scriptscriptstyle 2}}(x,p) :=
(2\pi\hbar) \{W_{\rho}(x,p)\}^2 \geq 0$. This quantity ${\mathcal
W}_{\rho^{\scriptscriptstyle 2}}(x,p)$ should be distinguished from
its counterpart $W_{\rho^{\scriptscriptstyle 2}}(x,p)$, which is
directly obtained from Eq. (\ref{eq:bopp_0-1}) with $\hat{\rho} \to
\hat{\rho}^2$ and thus may be negative valued like $W_{\rho}(x,p)$
itself. However, such formal symmetry between $\hat{\rho}^2$ and
${\mathcal W}_{\rho^{\scriptscriptstyle 2}}(x,p)$ is not available
for higher moments, for which $\mbox{Tr}(\hat{\rho}^l) \not\propto
\int dx dp\, \{W_{\rho}(x,p)\}^l$ with $l = 3,4,5,\cdots$
\cite{KIM18}. To directly discuss higher moments
$\mbox{Tr}(\hat{\rho}^{\alpha})$ with $\alpha \in {\mathbb R}^+$ in
the phase space, we therefore generalize Eq.
(\ref{eq:purity_measure1}) in such a way that the $(\alpha-1)$th
fractional moment of probability is given by
\begin{equation}\label{eq:wigner-entropy_1}
    \<p^{\alpha-1}\> = \mbox{Tr}(\hat{\rho}^{\alpha}) = \int dx\,dp\, {\mathcal W}_{\rho^{\scriptscriptstyle \alpha}}(x,p)\,,
\end{equation}
in which the quantity ${\mathcal W}_{\rho^{\scriptscriptstyle
\alpha}}(x,p) = (2\pi\hbar) \{W_{\rho^{{\scriptscriptstyle
\alpha/2}}}(x,p)\}^2 \geq 0$ correspondingly results from Eq.
(\ref{eq:bopp_0-1}) with $\hat{\rho} \to \hat{\rho}^{\alpha/2}$.
Here, a fractional operator $\hat{\rho}^{\alpha}$ is obtained from
the spectral expansion of $\hat{\rho}$ by substituting its
eigenvalues $p_n$'s with their positive $\alpha$th powers. We stress
that its phase-space counterpart $W_{\rho^{{\scriptscriptstyle
\alpha}}}(x,p)$, on the other hand, cannot directly be obtained from
$W_{\rho}(x,p)$.

Now we introduce non-negative distributions $\widetilde{\mathcal
W}_{\rho;\alpha}(x,p) := {\mathcal W}_{\rho^{\scriptscriptstyle
\alpha}}(x,p)/{\mathcal N}_{\rho^{\scriptscriptstyle \alpha}}$ with
the normalizing ${\mathcal N}_{\rho^{\scriptscriptstyle \alpha}} =
\int dx dp\, {\mathcal W}_{\rho^{\scriptscriptstyle \alpha}}(x,p) =
\mbox{Tr}(\hat{\rho}^{\alpha})$ such that $\int dx dp\,
\widetilde{\mathcal W}_{\rho;\alpha}(x,p) = 1$. Then, this
fractional moment may also be interpreted as the expectation value
${\mathcal N}_{\rho^{\scriptscriptstyle \alpha}} =
(\<\widetilde{\mathcal W}_{\scriptscriptstyle \rho;\alpha})^{-1}\,
{\mathcal W}_{\rho^{\scriptscriptstyle
\alpha}}\>_{\scriptscriptstyle \widetilde{\mathcal
W}_{\scriptscriptstyle \rho;\alpha}}$, expressed in terms of
$\<A\>_{\scriptscriptstyle \widetilde{\mathcal
W}_{\scriptscriptstyle \rho;\alpha}} := \int dx dp\,
A(x,p)\,\widetilde{\mathcal W}_{\scriptscriptstyle
\rho;\alpha}(x,p)$. For comparison only, we introduce other
distributions $\utilde{W}_{\rho;\alpha}(x,p) :=
W_{\rho^{\scriptscriptstyle
\alpha}}(x,p)/N_{\rho^{\scriptscriptstyle \alpha}}$, as well, with
$N_{\rho^{\scriptscriptstyle \alpha}} = \int dx dp\,
W_{\rho^{\scriptscriptstyle \alpha}}(x,p)$ such that $\int dx dp\,
\utilde{W}_{\rho;\alpha}(x,p) = 1$. Clearly, these distributions
$\utilde{W}_{\rho;\alpha}(x,p)$ can be negative-valued, though. Now,
let the set $\widetilde{\mathcal W}_{\rho} = \{\widetilde{\mathcal
W}_{\rho;\alpha'}(x,p)| \alpha'
> 0\}$, all elements of which correspond to the same density operator $\hat{\rho}$.
Then, the R\'{e}nyi entropy of $\hat{\rho}$ with order $\alpha$
($\in {\mathbb R}^+$) can be expressed as the R\'{e}nyi-Wigner
entropy of $\widetilde{\mathcal W}_{\rho}$ such that
\begin{equation}\label{eq:tilde_wigner1}
    S_{\alpha}(\widetilde{\mathcal W}_{\rho}) = (1-\alpha)^{-1} \ln\left({\mathcal N}_{\rho^{\scriptscriptstyle
    \alpha}}\right)\,,
\end{equation}
which is well-defined. Here, the particular selection of
$\widetilde{\mathcal W}_{\rho;\alpha}(x,p)$ with $\alpha' \to
\alpha$ from the set $\widetilde{\mathcal W}_{\rho}$ is required for
actual evaluation of ${\mathcal N}_{\rho^{\scriptscriptstyle
\alpha}}$ for a given order $\alpha$. This necessarily means that
$S_{\alpha}(\widetilde{\mathcal W}_{\rho}) \to
S_{\alpha}(\widetilde{\mathcal W}_{\rho;\alpha})$ indeed. Obviously,
this entropy $S_{\alpha}(\widetilde{\mathcal W}_{\rho})$ differs
from the expression $\overline{S}_{\alpha}(W_{\rho}) :=
(1-\alpha)^{-1} \ln\left[\int dx dp\, (2\pi\hbar)
\{W_{\rho}(x,p)\}^{\alpha}\right]$ for $\alpha \ne 2$; also note
Eqs. (\ref{eq:moment1})-(\ref{eq:moment4}). Eq.
(\ref{eq:tilde_wigner1}) is the first central result of our paper.
This result enables us to evaluate the R\'{e}nyi-$\alpha$ entropy in
the phase-space picture in a more compact way than its counterpart
provided in Ref. \cite{KIM18} (cf. Eq. (13) thereof) which has been
derived, on the other hand, for positive integers $\alpha \to l =
2,3,4,\cdots$ only and expressed in terms of the product of plain
Wigner functions $W_{\rho}(x,p)$ with the Bopp shift; as a result,
the normalizing ${\mathcal N}_{\rho^{\scriptscriptstyle \alpha}}$ in
Eq. (\ref{eq:tilde_wigner1}) now replaces the lengthy expression
\begin{equation}\label{eq:bopp_4-2}
    (2\pi\hbar)^{l-1} \int dx dp\, W_{\rho}(x,p)\,
    \left\{W_{\rho}\left(x-\frac{\hbar}{2i}\frac{\partial}{\partial p},
    p+\frac{\hbar}{2i}\frac{\partial}{\partial x}\right)\right\}^{l-2}\,
    W_{\rho}(x,p) = \mbox{Tr}(\hat{\rho}^l)\,.
\end{equation}
Then, we can compute the entropy with no need for an analytic
continuation of $l \to \alpha$ for arbitrary non-Gaussian states. In
fact, it is easy to expect that this analytic continuation will be a
formidable task for generic non-Gaussian states.

We will be interested especially in the case of $\alpha = 1$, for
which
\begin{equation}\label{eq:effective_pic1}
    \widetilde{\mathcal W}_{\rho} \to \widetilde{\mathcal W}_{\rho;1}(x,p) = {\mathcal W}_{\rho}(x,p) = (2\pi\hbar)\, \{W_{\rho^{{\scriptscriptstyle
    1/2}}}(x,p)\}^2 \geq 0\,,
\end{equation}
obviously with ${\mathcal W}_{\rho}(x,p) \ne W_{\rho}(x,p)$. Then,
the von-Neumann entropy simply reduces to
\begin{equation}\label{eq:wigner-entropy_2}
    S_1(\widetilde{\mathcal W}_{\rho}) = -\partial_{\alpha} \left.\ln\left({\mathcal N}_{\rho^{\scriptscriptstyle
    \alpha}}\right)\right|_{\alpha=1}\,,
\end{equation}
directly obtained without considering any analytic continuation at
all. We remark that the Shannon measure of information (or Shannon's
entropy), as the classical counterpart of the von-Neumann entropy,
will directly appear from Eq. (\ref{eq:wigner-entropy_2}) with
$\hbar \to 0$; in fact, it is known that all R\'{e}nyi entropies
$S_{\alpha}(\hat{\rho})$, and so $S_{\alpha}(\widetilde{\mathcal
W}_{\rho})$, tend asymptotically to the von-Neumann entropy in the
classical limit (e.g., \cite{LIN13,BRA15,PAT17}). It is also
worthwhile to point out that the ``Wigner entropy'' in the form of
$S_{\mbox{\tiny W}}(\widetilde{\mathcal W}_{\rho;\alpha}) := -\int
dx dp\, \widetilde{\mathcal
W}_{\rho;\alpha}(x,p)\,\ln\{(2\pi\hbar)\,\widetilde{\mathcal
W}_{\rho;\alpha}(x,p)\}$ is ill-defined, though $\widetilde{\mathcal
W}_{\rho;\alpha}(x,p) \geq 0$, because the Bopp shift has not been
employed at all and thus this expression, e.g., cannot appropriately
distinguish pure states from mixed states [cf. Eqs.
(\ref{eq:moment1})-(\ref{eq:moment4}) and the discussion
thereafter\,].

Now we examine some properties of a distribution
$\widetilde{\mathcal W}_{\rho;\alpha}(x,p)$. First, we point out
that its non-negative nature (like that of its classical
counterpart) cannot suitably reflect the orthogonality relation
between any pair $(n,m)$ of two different eigenstates, because of
the trace $\int dx dp\, \widetilde{\mathcal
W}_n(x,p)\,\widetilde{\mathcal W}_m(x,p)
> 0$. Therefore, this distribution cannot be
interpreted as a genuine probability distribution satisfying the
required quantum feature. For comparison, we simply point out that
this problem can be remedied finally with the help of the Bopp
shift, as in Eq. (\ref{eq:bopp_4-2}), such that, with
$\hat{\rho}^{\alpha} \to \hat{\rho}$ for pure states,
$\mbox{Tr}(\hat{\rho}_n \hat{\rho}_m)$ is expressed, instead, as
\begin{eqnarray}\label{eq:orthogonality_rel1}
    && (2\pi\hbar)^3 \int dx dp\, W_r(x,p)\,
    W_s\left(x-\frac{\hbar}{2i}\frac{\partial}{\partial p},
    p+\frac{\hbar}{2i}\frac{\partial}{\partial x}\right)\, W_u\left(x-\frac{\hbar}{2i}\frac{\partial}{\partial p},
    p+\frac{\hbar}{2i}\frac{\partial}{\partial x}\right)\, W_v(x,p)\n\\
    &&= \delta_{nm}
\end{eqnarray}
indeed, in which $(r,s,u,v) = (n,n,m,m); (n,m,m,n); (m,m,n,n);
(m,n,n,m)$, thus equivalent to $\mbox{Tr}\{(\hat{\rho}_n)^2
(\hat{\rho}_m)^2\} = \delta_{nm}$. In the limit of $\hbar \to 0$,
the left-hand side of Eq. (\ref{eq:orthogonality_rel1}) would simply
reduce to $\int dx dp\, \widetilde{\mathcal
W}_n(x,p)\,\widetilde{\mathcal W}_m(x,p) > 0$ and therefore the
orthogonality relation would be gone completely, which is exactly
the case for any classical probability distribution $P(x,p)$.

Second, let us consider the expectation value of an observable
$\hat{A}$ within our formulation. For the sake of simplicity, we now
restrict our discussion into the case of $\alpha = 1$. It is then
straightforward to show that $\mbox{Tr}(\hat{\rho} \hat{A}) \ne \int
dx dp\, \widetilde{W}_{\rho;1}(x,p)\, W_{\scriptscriptstyle
A}(x.p)$, but the expression
\begin{equation}\label{eq:expectation-value-wrtW1}
    \mbox{Tr}(\hat{\rho} \hat{A}) = \mbox{Tr}(\hat{\rho}^{\scriptscriptstyle 1/2} \hat{A}\, \hat{\rho}^{\scriptscriptstyle 1/2}) =
    (2\pi\hbar) \int dx dp\, W_{\rho^{\scriptscriptstyle
    1/2}}(x,p)\, W_{\scriptscriptstyle A}\left(x-\frac{\hbar}{2i} \frac{\partial}{\partial p},
    p+\frac{\hbar}{2i} \frac{\partial}{\partial x}\right)\; W_{\rho^{\scriptscriptstyle
    1/2}}(x,p)\,.
\end{equation}
This result is reminiscent of the expression $\<\hat{A}\>_{\psi} =
\int dx\, \psi^{\ast}(x)\, A(x, -i\hbar\,\partial_x)\, \psi(x)$ in
formal similarity. Subsequently, with the help of the Fourier
transform ${\mathfrak W}_{\scriptscriptstyle A}({\mathfrak
x},{\mathfrak p}) = (2\pi\hbar)^{-1} \int dx dp\,
W_{\scriptscriptstyle A}(x,p)\, \exp\{i\,(x {\mathfrak p} + p
{\mathfrak x})/\hbar\}$, Eq. (\ref{eq:expectation-value-wrtW1}) will
finally be transformed, after some algebraic manipulations, into
\begin{eqnarray}\label{eq:expectation-value-wrtW1-1}
    &&\mbox{Tr}(\hat{\rho} \hat{A}) = (4\pi\hbar)^{-1} \int_{-\infty}^{\infty}
    dx\,dp\,dx_1\,dp_1\,dx_2\,dp_2\;
    W_{\rho^{\scriptscriptstyle 1/2}}(x/2,p)\, \exp\left\{-\frac{i}{\hbar}\,(x_1-x_2)\,p\right\}\; \times\n\\
    &&W_{\scriptscriptstyle A}(x_1/2,p_1)\,
    \exp\left\{-\frac{i}{\hbar}\,(x_2-x)\,p_1\right\}\;
    W_{\rho^{\scriptscriptstyle 1/2}}(x_2/2,p_2)\,\exp\left\{-\frac{i}{\hbar}\,(x-x_1)\,p_2\right\}\,.
\end{eqnarray}
Considering in Eq. (\ref{eq:expectation-value-wrtW1-1}) the
(diagonal) terms with $x = x_1 = x_2$ and $p = p_1 = p_2$ only, then
Eq. (\ref{eq:expectation-value-wrtW1-1}) would reduce to the compact
form given by $\int dx dp\, \widetilde{W}_{\rho;1}(x,p)\,
W_{\scriptscriptstyle A}(x.p)$, which is obviously not tantamount to
$\mbox{Tr}(\hat{\rho} \hat{A})$ for a generic observable $\hat{A}
\ne \id$. This also shows that the distribution
$\widetilde{W}_{\rho;1}(x,p)$ has the conceptual drawback that this
quantity cannot be interpreted as a quasi-probability distribution
over phase space, either. However, it is easy to show that for
$\hat{A} \to \hat{\rho}^{\alpha-1}$ leading to $[\hat{A},
\hat{\rho}^{1/2}] = 0$, Eqs. (\ref{eq:expectation-value-wrtW1}) and
(\ref{eq:expectation-value-wrtW1-1}) exactly reduce to Eq.
(\ref{eq:wigner-entropy_1}). Therefore, the distribution
$\widetilde{\mathcal W}_{\rho;\alpha}(x,p)$, despite its conceptual
drawback discussed above, is still useful (in the operational sense)
for evaluations of the R\'{e}nyi-$\alpha$ entropies in the
phase-space picture.

From the discussions provided in the preceding paragraphs, we arrive
at the following conclusion concerning our central result: The usual
Wigner function $W_{\rho}(x,p)$ of a non-Gaussian state $\hat{\rho}$
(as well as any product of the Wigner functions) cannot produce its
R\'{e}nyi-$\alpha$ entropies (with $\alpha \in {\mathbb R}^+$). On
the other hand, the non-negative distribution $\widetilde{\mathcal
W}_{\rho;\alpha}(x,p)$ cannot suitably be interpreted as a
(quasi)-probability distribution, but it is actually this
distribution that can produce the R\'{e}nyi entropies $S_{\alpha}$
in the phase-space picture. Therefore, it is legitimate to say that
this distribution is an operational one for exact evaluation of the
R\'{e}nyi entropies and thus for systematic access to the
higher-moment information (including $S_1$) of a non-Gaussian state
$\hat{\rho}$; again, without ${\mathcal W}_{\rho^{\scriptscriptstyle
\alpha}}(x,p)$, the analytic continuation for order $\alpha$ would
necessarily be required for evaluation of, e.g., $S_1$, as discussed
already, which will however be a formidable task for generic
non-Gaussian states. As a result, we may also claim that while the
usual Wigner function $W_{\rho}(x,p)$, as a well-defined
quasi-probability distribution, enables us to compute the
expectation value of an observable via Eqs. (\ref{eq:expectation1})
and (\ref{eq:bopp_0-2}), it cannot cover the full information
available in a given density operator $\hat{\rho}$, without, e.g.,
the supplemental quantities ${\mathcal W}_{\rho^{\scriptscriptstyle
\alpha}}(x,p)$ (or $W_{\rho^{\scriptscriptstyle \alpha}}(x,p)$).

It is also tempting to ask about difference between the two
fractional quantities, ${\mathcal W}_{\rho^{\scriptscriptstyle
\alpha}}(x,p)$ and $W_{\rho^{\scriptscriptstyle \alpha}}(x,p)$, in
addition to whether they are non-negative or not. To do so, let us
rewrite the fractional moment given in Eq.
(\ref{eq:wigner-entropy_1}) as
\begin{eqnarray}\label{eq:purity_rel1}
    \mbox{Tr}(\hat{\rho}^{\alpha}) &=& (2\pi\hbar)^3\,({\mathcal
    N}_{\rho^{\scriptscriptstyle \alpha}})^{-2} \int dx dp\; W_{\rho^{\scriptscriptstyle
    \alpha/2}}(x,p)\; W_{\rho^{\scriptscriptstyle \alpha/2}}\left(x-\frac{\hbar}{2i}\frac{\partial}{\partial p},
    p+\frac{\hbar}{2i}\frac{\partial}{\partial x}\right) \times\n\\
    && W_{\rho^{\scriptscriptstyle \alpha/2}}\left(x-\frac{\hbar}{2i}\frac{\partial}{\partial p},
    p+\frac{\hbar}{2i}\frac{\partial}{\partial x}\right)\; W_{\rho^{\scriptscriptstyle
    \alpha/2}}(x,p)\,.
\end{eqnarray}
Then, we apply the same technique as for Eq.
(\ref{eq:expectation-value-wrtW1-1}) with the help of the Fourier
transform ${\mathfrak W}_{\rho^{\scriptscriptstyle
\alpha}}({\mathfrak x},{\mathfrak p}) = (2\pi\hbar)^{-1} \int dx
dp\, W_{\rho^{\scriptscriptstyle \alpha}}(x,p)\, \exp\{i\,(x
{\mathfrak p} + p {\mathfrak x})/\hbar\}$, which will enable Eq.
(\ref{eq:purity_rel1}) to have the form
\begin{eqnarray}\label{eq:purity_rel1-1}
    &&\mbox{Tr}(\hat{\rho}^{\alpha}) = ({\mathcal
    N}_{\rho^{\scriptscriptstyle \alpha}})^{-2} \int_{-\infty}^{\infty} dx\,dp\,dx_1\,dp_1\,dx_2\,dp_2\,dx_3\,dp_3\, \times\\
    &&W_{\rho^{\scriptscriptstyle \alpha/2}}(x,p)\,\exp\left\{-\frac{i}{\hbar}\,(x_1+x_2+x_3)\,p\right\}\, W_{\rho^{\scriptscriptstyle \alpha/2}}(x +
    (x_2+x_3)/2,p_1)\,\exp\left(\frac{i}{\hbar}\,x_1\,p_1\right)\, \times\n\\
    &&W_{\rho^{\scriptscriptstyle \alpha/2}}(x -
    (x_1-x_3)/2,p_2)\,\exp\left(\frac{i}{\hbar}\,x_2\,p_2\right)\, W_{\rho^{\scriptscriptstyle \alpha/2}}(x -
    (x_1+x_2)/2,p_3)\,\exp\left(\frac{i}{\hbar}\,x_3\,p_3\right)\,.\n
\end{eqnarray}
As observed, this expression, being not in terms of our non-negative
quantity ${\mathcal W}_{\rho^{\scriptscriptstyle \alpha}}(x,p)$, has
higher computational complexity (as its drawback) than its
counterpart, Eq. (\ref{eq:wigner-entropy_1}), even for integer
orders $\alpha \to l$ (cf. Eq. (\ref{eq:purity_rel1-1-neu}) for an
actual evaluation of Eq. (\ref{eq:purity_rel1-1}) with respect to a
particular state).

Now, we consider the canonical thermal equilibrium state
$\hat{\rho}_{\scriptscriptstyle \beta}$ for explicit evaluation of
$\widetilde{W}_{\rho;\alpha}(x,p)$. First, its Wigner function is
given by $W_{\beta}(x,p) = \{(2\pi\hbar)\,Z_{\beta}\}^{-1}\,
\mbox{Num}(\beta)$, where the partition function $Z_{\beta}$ and the
numerator
\begin{equation}\label{eq:Wigner_thermal-0}
    \mbox{Num}(\beta) := (2\pi\hbar)\, \sum_n
     \exp\left(-\beta E_n\right)\; W_n(x,p)\,.
\end{equation}
By noting that $\mbox{Tr}\{(\hat{\rho}_{\scriptscriptstyle
\beta})^{\alpha}\} = Z_{\alpha\beta}/(Z_{\beta})^{\alpha}$, it is
straightforward to show that $\widetilde{\mathcal
W}_{\rho;\alpha}(x,p) =: \widetilde{\mathcal W}_{\beta;\alpha}(x,p)
= (2\pi\hbar)\,\{W_{\rho^{\scriptscriptstyle
\alpha/2}}(x,p)\}^2\,\{\mbox{Tr}(\rho^{\alpha})\}^{-1} =
\{(2\pi\hbar)\,Z_{\alpha\beta}\}^{-1}\,
\{\mbox{Num}(\alpha\beta/2)\}^2 \geq 0$, where
$W_{\rho^{\scriptscriptstyle \alpha}}(x,p) =
(Z_{\beta})^{-\alpha}\,(Z_{\alpha\beta})\,W_{\alpha\beta}(x,p)$.
Therefore, $\widetilde{\mathcal W}_{\beta;\alpha}(x,p)$ can be
obtained from $W_{\beta}(x,p)$ simply by substitution of both
$Z_{\beta} \to Z_{\alpha\beta}$ and $\mbox{Num}(\beta) \to
\{\mbox{Num}(\alpha\beta/2)\}^2$, which is actually valid for the
canonical thermal state of an arbitrary quantum system. As a simple
example, the thermal state of a single linear oscillator, being
Gaussian, is considered such that $Z_{\beta} =
2^{-1}\,\mbox{csch}(\beta\hbar\omega/2)$ \cite{ING02}, and
\begin{subequations}
\begin{eqnarray}
    W_{\beta}(x,p) &=& \frac{\mbox{sech}(\beta\hbar\omega/2)}{(2\pi\hbar)\,Z_{\beta}}\,
    \exp\left[-\left(\tanh \frac{\beta\hbar\omega}{2}\right) \left\{(\kappa x)^2 + \frac{p^2}{(\hbar
    \kappa)^2}\right\}\right]\, \geq\, 0\label{eq:wigner-oscillator1}\\
    \widetilde{\mathcal W}_{\beta;1}(x,p) &=&
    \frac{\{\mbox{sech}(\beta\hbar\omega/4)\}^2}{(2\pi\hbar)\,Z_{\beta}}\,
    \exp\left[-2 \left(\tanh \frac{\beta\hbar\omega}{4}\right) \left\{(\kappa x)^2 + \frac{p^2}{(\hbar
    \kappa)^2}\right\}\right]\, \geq\, 0\label{eq:wigner-oscillator1-1}
\end{eqnarray}
\end{subequations}
where $\kappa = (m \omega /\hbar)^{1/2}$, as well as the Wigner
entropy $S_{\mbox{\tiny W}}(W_{\beta}) = S_2(W_{\beta}) + 1 - \ln(2)
\ne S_1(W_{\beta})$. It is then straightforward to verify Eqs.
(\ref{eq:tilde_wigner1}) and (\ref{eq:wigner-entropy_2}) for this
system. We also note that in the limit of $\hbar \to 0$ leading to
$(2\pi\hbar)\,Z_{\beta} \to (Z_{\beta,\mbox{\sc cl}})$, Eqs.
(\ref{eq:wigner-oscillator1}) and (\ref{eq:wigner-oscillator1-1})
will reduce to their classical counterpart $P_{\beta}(x,p) =
(Z_{\beta,\mbox{\sc cl}})^{-1}\,e^{-\beta\,(p^2/2m\,+\,m \omega^2
x^2/2)}$.

Finally, we stress that our formulation for the study of
non-Gaussian states, consisting of the non-negative phase-space
distributions and R\'{e}nyi-Wigner entropies, essentially differs
from the approach based on the (non-negative) Husimi functions
$Q_{\rho}(x,p) = \<\gamma|\hat{\rho}|\gamma\>/\pi$, defined in terms
of the coherent state $|\gamma\>$ with $\gamma = 2^{-1/2}\,(\kappa x
+ i p/\hbar\kappa)$, and the resulting Wehrl entropies defined as
$(-1) \int dq dp\; Q_{\rho}(x,p)\,\ln\{Q_{\rho}(x,p)\}$; this
entropy has been known to have a conceptual weakness that results
from the non-orthogonality $|\<\gamma_1|\gamma_2\>|^2 =
e^{-|\gamma_1-\gamma_2|^2}$, where $|\gamma_1\>$ and $\gamma_2\>$
denote different coherent states (e.g., \cite{GNU01,KIM18}).

\section{Wigner function of thermal state for one-dimensional box problem}\label{sec:sec3}
%
The system under consideration is a single particle confined in the
region of $-a \leq x \leq a$ (with $a > 0$) by a one-dimensional
infinite potential well with either the Dirichlet boundary condition
$\Psi(a) = \Psi(-a) = 0$ (Dbc) or Neumann boundary condition
$\Phi'(a) = \Phi'(-a) = 0$ (Nbc). As is well-known, its $n$th
eigenstate for Dbc is given by \cite{LEE83,ALM90,GRO94,LEE95,DIA02}
\begin{subequations}
\begin{equation}\label{eq:box-probblem-nth-eigen-di}
    \psi_n(x) = \left(\frac{1}{a}\right)^{1/2}\, \sin\left\{\frac{n\pi}{2}\left(1 + \frac{x}{a}\right)\right\}\,,
\end{equation}
where $n = 1, 2, 3, \cdots$ and $|x| \leq a$, while the eigenvalue
for Nbc is given by
\begin{equation}\label{eq:box-probblem-nth-eigen}
    \phi_n(x) = \left(\frac{1}{a}\right)^{1/2}\, \cos\left\{\frac{n\pi}{2}\left(1 + \frac{x}{a}\right)\right\}\,,
\end{equation}
\end{subequations}
where $n = 1, 2, 3, \cdots$ and $|x| \leq a$, and
$\phi_{\scriptscriptstyle 0}(x) = (2a)^{-1/2}$; therefore,
$\phi_n(x)$ is discontinuous at $x = \pm a$ if the analytic
continuation is under consideration that $\phi_n(x) \equiv 0$ for
$|x| > a$. The corresponding energy eigenvalue is $E_n(L) =
(p_n)^2/(2m)$ for both Dbc and Nbc, where $m$ is the mass of the
particle, and $p_n = \hbar k_n$ with $k_n = \pm n\pi/L$; here $L =
2a$ denotes the width of the potential well (note that $E_0 = 0$ for
Nbc \cite{FAC15}). Then, it is straightforward to compute the Wigner
function corresponding to the eigenstate $|n\>$ such that for Dbc,
\begin{subequations}
\begin{eqnarray}
    \hspace*{-1cm}W_{n;D}(x,p) &=& \frac{1}{\pi\hbar} \int_{-\xi_x}^{\xi_x}
    d\xi\; \psi_n(x+\xi)\; \psi_n(x-\xi)\; e^{-2 i p \xi/\hbar}\label{eq:wigner_box_1-0}\n\\
    \hspace*{-1cm}&=& \frac{1}{4\pi a} \left[\left\{\sin\left(\frac{2}{\hbar}\,\xi_x\,(p + p_n)\right)\times\left(\frac{1}{p+p_n} - \frac{1}{p}\right)\right\} +
    \{p_n \to -p_n\}\right]\label{eq:wigner_box_1}
\end{eqnarray}
(cf. \cite{LEE83,ALM90,LEE95,DIA02}) while for Nbc,
\begin{eqnarray}
    \hspace*{-1cm}W_{n;N}(x,p) &=& \frac{1}{\pi\hbar} \int_{-\xi_x}^{\xi_x}
    d\xi\; \phi_n(x+\xi)\; \phi_n(x-\xi)\; e^{-2 i p \xi/\hbar}\label{eq:wigner_box_1-n}\n\\
    \hspace*{-1cm}&=& \frac{1}{4\pi a} \left[\left\{\sin\left(\frac{2}{\hbar}\,\xi_x\,(p + p_n)\right)\times\left(\frac{1}{p+p_n} + \frac{1}{p}\right)\right\} +
    \{p_n \to -p_n\}\right]\label{eq:wigner_box_n}
\end{eqnarray}
\end{subequations}
where $n = 1,2,3,\cdots$ and $W_{\scriptscriptstyle 0;N}(x,p) =
(2\pi a p)^{-1} \sin(2 \xi_x p/\hbar)$; here, $\xi_x = a-|x|$ is
required by the boundary condition of $|x + \xi| \leq a$ and $|x-
\xi| \leq a$. Therefore, $W_n(\pm a,p) = 0$ for both Dbc and Nbc, as
required (cf. Ref. \cite{DIA02}). We observe that Eqs.
(\ref{eq:wigner_box_1}) and (\ref{eq:wigner_box_n}), as well as
$W_{\scriptscriptstyle 0;N}(x,p)$, are even functions of both $x$
and $p$, non-Gaussian, and can be negative valued indeed (cf. Figs.
\ref{fig:fig1} and \ref{fig:fig2}).

It is also instructive, for later purposes, to consider Eqs.
(\ref{eq:wigner_box_1}) and (\ref{eq:wigner_box_n}) in the limit of
$\hbar \to 0$, in particular the respective ground states. First,
with the help of the identity $\delta(p) = \lim_{\epsilon\to 0}\,
(\pi p)^{-1}\, \sin(p/\epsilon)$ \cite{GAS99}, it is easy to show
that $W_{0;N}(x,p) \to (2a)^{-1}\,\delta(p)$ in this limit. It is
also straightforward to observe that $W_{1;D}(x,p) \to 0$ for $p \ne
0$ and $W_{1;D}(x,0) \to \infty$, thus yielding $W_{0;D}(x,p) \to
(2a)^{-1}\,\delta(p)$ as well.

Now, we are ready to discuss the thermal Wigner function of this
system, which is given for Dbc and Nbc by [cf. Eq.
(\ref{eq:Wigner_thermal-0})]
\begin{subequations}
\begin{eqnarray}
     W_{\beta;D}(x,p) &=& \frac{1}{(Z_{\beta})_{\scriptscriptstyle D}} \sum_{n=1}^{\infty}
     \exp\left(-\lambda n^2\right)\;
     W_{n;D}(x,p)\,,\label{eq:Wigner_thermal-1}\\
     W_{\beta;N}(x,p) &=& \frac{1}{(Z_{\beta})_{\scriptscriptstyle N}} \sum_{n=0}^{\infty}
     \exp\left(-\lambda n^2\right)\;
     W_{n;N}(x,p)\label{eq:Wigner_thermal-N}
\end{eqnarray}
\end{subequations}
with $\lambda = \beta\hbar^2\pi^2\,(8 m a^2)^{-1}$ and
$(Z_{\beta})_{\scriptscriptstyle N} =
(Z_{\beta})_{\scriptscriptstyle D} + 1$, respectively. These can
also be expressed as the integral form
\begin{eqnarray}\label{eq:jacobi-fkts2}
    W_{\beta}(x,p) &=& \frac{1}{(2\pi\hbar)\,a\,Z_{\beta}}
    \left\{B \left(\frac{\hbar}{2p}\right)\; \sin\left(\frac{2 \xi_x
    p}{\hbar}\right)\;
    \vartheta_4\left(\frac{\pi x}{2a},\exp\left(-\lambda\right)\right) +\right.\n\\
    && \left.\int_0^{\xi_x} d\xi\; \cos\left(\frac{2 \xi
    p}{\hbar}\right)\;
    \vartheta_3\left(\frac{\pi \xi}{2a},\exp\left(-\lambda\right)\right)\right\}
\end{eqnarray}
in terms of the Jacobi theta functions \cite{GRA07}
\begin{equation}
    \vartheta_3(z,q) = 1 + 2 \sum_{n=1}^{\infty}
    q^{n^2}\,\cos(2nz)\;\; ;\;\;
    \vartheta_4(z,q) = 1 + 2 \sum_{n=1}^{\infty}
    (-1)^n\,q^{n^2}\,\cos(2nz)\,;
\end{equation}
here, $Z_{\beta} \to (Z_{\beta})_{\scriptscriptstyle D} =
2^{-1}\,\{\vartheta_3(0,e^{-\lambda}) - 1\}$ and $B = -1$ for
$W_{\beta;D}(x,p)$ while $Z_{\beta} \to
(Z_{\beta})_{\scriptscriptstyle N} =
2^{-1}\,\{\vartheta_3(0,e^{-\lambda}) + 1\}$ and $B = 1$ for
$W_{\beta;N}(x,p)$.

To study the quantum-classical transition, we intend to rewrite Eqs.
(\ref{eq:Wigner_thermal-1}) and (\ref{eq:Wigner_thermal-N}); after
some algebraic manipulations, every single step of which is provided
in detail in the Appendix, we can finally arrive at the expression
\begin{eqnarray}\label{eq:wigner-thermal_1-3}
    \hspace*{-.0cm}W_{\beta}(x,p) &=& \frac{1}{(2\pi\hbar)\,Z_{\beta}}
    \left[\exp\left(-\frac{\beta p^2}{2m}\right)\sum_{\nu=-\infty}^{\infty}
    \mbox{Re}\left\{\exp\left(\frac{2 i p\,\nu L}{\hbar}\right)
    \left[\mbox{erf}\left(\left(\frac{2m}{\beta \hbar^2}\right)^{\scriptscriptstyle 1/2}\left(\xi_x + \nu L\right)\right.\right.\right.\right.\n\\
    \hspace*{-.0cm}&&+\left.\left.\left.\left.i p\left(\frac{\beta}{2m}\right)^{\scriptscriptstyle 1/2}\right) -
    \mbox{erf}\left(\left(\frac{2m}{\beta \hbar^2}\right)^{\scriptscriptstyle 1/2} \nu L + i
    p\left(\frac{\beta}{2m}\right)^{\scriptscriptstyle 1/2}\right)\right]\right\}\right.\n\\
    \hspace*{-.0cm}&&+\left.\frac{B Z_{\beta,\mbox{\sc cl}}}{2\pi a p}\,\sin\left(\frac{2 \xi_x
    p}{\hbar}\right)
    \sum_{\mu=-\infty}^{\infty}
    \exp\left\{-\frac{2m}{\beta \hbar^2}\left(\xi_x + \mu L\right)^2\right\}\right]
\end{eqnarray}
in terms of the error function $\mbox{erf}(z)$, where the width $L =
2a$ and the classical partition function $Z_{\beta,\mbox{\sc cl}} =
(8 \pi m a^2/\beta)^{1/2}$ for both Dbc and Nbc. Then, in the
classical limit, Eq. (\ref{eq:wigner-thermal_1-3}) reduces to its
classical counterpart $P_{\beta}(x,p) = (Z_{\beta,\mbox{\sc
cl}})^{-1}\,e^{-\beta p^2/2m}
> 0$, being Gaussian, which results from the term of $\nu = 0$ (with
$\hbar \to 0$). Eq. (\ref{eq:wigner-thermal_1-3}) is the second
central result of our paper.

Comments are deserved here. First, we observe that $W_{\beta}(\pm
a,p) = 0$ and thus $W_{\beta}(x,p)$ is continuous in the entire
phase space. On the other hand, $P_{\beta}(\pm a,p) \ne 0$ and thus
$P_{\beta}(x,p)$ is discontinuous at both boundary points. This
discontinuity also implies disappearance of the wave properties.
Second, the classical probability distribution $P_{\beta}(x,p)$
further reduces to $(2a)^{-1}\,\delta(p)$ at $T = 0$, in accordance
with both $W_{1;D}(x,p)$ and $W_{0;N}(x,p)$ within $\hbar \to 0$, as
discussed after Eq. (\ref{eq:wigner_box_n}). Third, all other terms
of $\nu \ne 0$ of Eq. (\ref{eq:wigner-thermal_1-3}) will then
represent the purely quantum correction; the value $2L$ denotes the
length of an arbitrary {\em primitive} periodic orbit (i.e., a
closed path traversed only {\em once} from an arbitrary phase-space
position $(x_0,p_0)$ to the same one after two reflections on the
potential walls at $x = \pm a$) \cite{KEA87,BRA97,STE98}. In fact,
if an index $\nu=\nu_{\mbox{\sc e}}$ (or $\mu=\mu_{\mbox{\sc e}}$)
is even, then it represents a periodic orbit with its length
$\nu_{\mbox{\sc e}} L$ (or $\mu_{\mbox{\sc e}} L$), corresponding to
$\nu_{\mbox{\sc e}}/2$ (or $\mu_{\mbox{\sc e}}/2$) repetitions of
its primitive periodic orbit. On the other hand, if an index
$\nu=\nu_{\mbox{\sc o}}$ (or $\mu=\mu_{\mbox{\sc o}}$) is odd, then
it represents an orbit moving from $(x_0,p_0)$ to $(-x_0,-p_0)$ with
its length $\nu_{\mbox{\sc o}} L$ (or $\mu_{\mbox{\sc o}} L$), which
is also needed due to the even parity of this system; note that the
cases of $\nu, \mu < 0$ simply denote periodic orbits initially
moving in the negative direction.

To explicitly discuss Eq. (\ref{eq:wigner-thermal_1-3}) in the
high-temperature regime, we employ both identities $\mbox{erf} = 1 -
\mbox{erfc}$ and $e^{2 z_1 z_2 + z_2^2}\,\mbox{erfc}(z_1 + z_2) =
\sum_{k=0}^{\infty} (-2 z_2)^k\, i^k \mbox{erfc}(z_1)$ \cite{ABR65},
which will yield the exact expression
\begin{eqnarray}\label{eq:orbits-temp1}
    \hspace*{-.0cm}&&W_{\beta}(x,p) = \frac{1}{(2\pi\hbar)\,Z_{\beta}}
    \left[\exp\left(-\frac{\beta p^2}{2m}\right) - \cos\left(\frac{2 \xi_x
    p}{\hbar}\right) \sum_{k=0}^{\infty} \left(-\frac{2\beta p^2}{m}\right)^k \times\right.\n\\
    && \sum_{\nu=1}^{\infty}
    \left\{i^{2k}\mbox{erfc}\left\{\left(\frac{2m}{\beta \hbar^2}\right)^{\scriptscriptstyle 1/2}\left(\xi_x + (\nu-1) L\right)\right\} -
    i^{2k}\mbox{erfc}\left\{\left(\frac{2m}{\beta \hbar^2}\right)^{\scriptscriptstyle 1/2}\left(\nu L-\xi_x\right)\right\}\right\}\,-\n\\
    \hspace*{-.0cm}&&\frac{Z_{\beta,\mbox{\sc cl}}}{4 \pi^{\scriptscriptstyle 1/2}\,a p}\,\sin\left(\frac{2 \xi_x
    p}{\hbar}\right) \sum_{k=1}^{\infty} \left(-\frac{2\beta p^2}{m}\right)^k \times\\
    && \sum_{\nu=1}^{\infty}
    \left\{i^{2k-1}\mbox{erfc}\left\{\left(\frac{2m}{\beta \hbar^2}\right)^{\scriptscriptstyle 1/2}\left(\xi_x + (\nu-1) L\right)\right\} +
    i^{2k-1}\mbox{erfc}\left\{\left(\frac{2m}{\beta \hbar^2}\right)^{\scriptscriptstyle 1/2}\left(\nu L-\xi_x\right)\right\}\right\}\,+\n\\
    \hspace*{-.0cm}&&\left.\frac{B Z_{\beta,\mbox{\sc cl}}}{2\pi a p}\,\sin\left(\frac{2 \xi_x
    p}{\hbar}\right)
    \sum_{\mu=1}^{\infty} \left\{\exp\left\{-\frac{2m}{\beta \hbar^2}\left(\xi_x + (\mu-1) L\right)^2\right\} +
    \exp\left\{-\frac{2m}{\beta \hbar^2}\left(\mu
    L-\xi_x\right)^2\right\}\right\}\right]\,.\n
\end{eqnarray}
With the help of $i^k\mbox{erfc}(0) = \{2^k\,\Gamma(k/2 + 1)\}^{-1}$
with $i^{-1}\mbox{erfc}(z) = (2/\sqrt{\pi}) \exp(-z^2)$
\cite{ABR65}, the boundary condition $W_{\beta}(\pm a,p) = 0$ can be
confirmed. We note here that the classical Gaussian part and the
quantal non-Gaussian part compete with each other, which is not the
case for a single linear oscillator, Eq.
(\ref{eq:wigner-oscillator1}). This non-Gaussian part is actually
expressed as two different kinds of contributions; the sums over the
periodic orbits $(\nu, \mu)$ are responsible for the purely quantum
effect (i.e., temperature-independent) while the sums of $k$ for the
thermal effect (also note, for comparison, that $\hbar$ and $\beta$
are always non-separable in form of $\beta\hbar\omega$ for Eq.
(\ref{eq:wigner-oscillator1})). As is well-known, the sums over
periodic orbits with non-zero lengths are responsible for the
stepwise nature of the spectral staircase $N(E) = \sum_n \Theta(E -
E_n)$ while the trivial orbits with zero lengths solely contribute
to the smooth increase of $N(E)$ with $E$ \cite{KEA87,BRA97,STE98}.

Therefore, it is interesting to consider two different limits of Eq.
(\ref{eq:orbits-temp1}) separately; first, the purely semiclassical
limit, by neglecting all periodic orbits with non-zero lengths
(i.e., weakening the oscillatory quantum correction), and second,
the high-temperature limit ($\beta \to 0$). First, in the
semiclassical limit, Eq. (\ref{eq:orbits-temp1}) easily reduces to
\begin{eqnarray}\label{eq:semiclassical_wigner-thermal_1-1-1}
    \hspace*{-.0cm}W_{\beta}(x,p) &\approx& \frac{1}{\overline{N}_{\beta}} \left[\exp\left(-\frac{\beta p^2}{2m}\right) - \cos\left(\frac{2 \xi_x
    p}{\hbar}\right) \sum_{k=0}^{\infty} \left(-\frac{2\beta p^2}{m}\right)^k
    i^{2k}\mbox{erfc}\left\{\left(\frac{2m}{\beta \hbar^2}\right)^{\scriptscriptstyle 1/2} \xi_x\right\}\, -\right.\n\\
    \hspace*{-.0cm}&& \left(\frac{m}{2\beta}\right)^{\scriptscriptstyle 1/2} \frac{1}{p}\,\sin\left(\frac{2 \xi_x
    p}{\hbar}\right) \sum_{k=1}^{\infty} \left(-\frac{2\beta p^2}{m}\right)^k
    i^{2k-1}\mbox{erfc}\left\{\left(\frac{2m}{\beta \hbar^2}\right)^{\scriptscriptstyle 1/2} \xi_x\right\}\,+\n\\
    \hspace*{-.0cm}&& \left.\left(\frac{2m}{\pi\beta}\right)^{\scriptscriptstyle 1/2} \frac{B}{p}\,\sin\left(\frac{2 \xi_x
    p}{\hbar}\right) \exp\left\{-\frac{2m}{\beta \hbar^2} \left(\xi_x\right)^2\right\}\right]
\end{eqnarray}
with the corresponding normalizing $\overline{N}_{\beta}$. Eq.
(\ref{eq:semiclassical_wigner-thermal_1-1-1}) actually meets the
boundary condition $W_{\beta}(\pm a, p) = 0$, as long as $\hbar$ is
finite, albeit sufficiently small. Then, Figs. \ref{fig:fig3} and
\ref{fig:fig4} show that even this semiclassical result with the
weakened oscillatory quantum correction can possess negative values
indeed. On the other hand, in the high-temperature limit, Eq.
(\ref{eq:orbits-temp1}) turns out to be
\begin{equation}\label{eq:high-temp-regime_1}
    W_{\beta}(x,p) \stackrel{\beta \to 0}{=} \frac{1}{Z_{\beta,\mbox{\sc cl}}}
    \left[\exp\left(-\frac{\beta p^2}{2m}\right) + \mbox{QF}_{\beta}(x,p)\right]\,,
\end{equation}
where the quantum fluctuation
\begin{eqnarray}
    \mbox{QF}_{\beta}(x,p) &=& \frac{B}{(\pi\beta/2m)^{\scriptscriptstyle 1/2}\,p}\, \left\{1 + \mathcal O(\beta)\right\}\, \sin\left(\frac{2 \xi_x
    p}{\hbar}\right) \times\label{eq:quantum-f1}\\
    \hspace*{-.0cm}&&\sum_{\mu=1}^{\infty} \left[\exp\left\{-\frac{2m}{\beta
    \hbar^2}\left(\xi_x + (\mu-1) L\right)^2\right\} +
    \exp\left\{-\frac{2m}{\beta \hbar^2}\left(\mu
    L-\xi_x\right)^2\right\}\right]\n
\end{eqnarray}
with $\int dx dp\; \mbox{QF}_{\beta}(x,p) = 0$. Eq.
(\ref{eq:high-temp-regime_1}) also can be negative valued, as long
as $\beta$ is finite, albeit sufficiently small. This cannot meet
the boundary condition (cf. Fig. \ref{fig:fig5}).

Now, the non-negative distribution $\widetilde{\mathcal
W}_{\beta;\alpha}(x,p)$ for the canonical thermal state can directly
be obtained from the usual Wigner function $W_{\beta}(x,p)$ by
utilizing the scenario for the canonical thermal state, discussed
after Eq. (\ref{eq:Wigner_thermal-0}). Therefore, it is
straightforward to have
\begin{subequations}
\begin{eqnarray}
    \widetilde{\mathcal W}_{\beta;\alpha}(x,p) &=& (2\pi\hbar)\,
    (Z_{\scriptscriptstyle \alpha\beta/2})^2\, \{W_{\scriptscriptstyle \alpha\beta/2}(x,p)\}^2 /Z_{\alpha\beta}\\
    S_{\alpha}(\widetilde{\mathcal W}_{\beta}) &=& (1 - \alpha)^{-1}\,
    \ln\left[(2\pi\hbar)\,(Z_{\beta})^{-\alpha}\,(Z_{\scriptscriptstyle \alpha\beta/2})^2 \int dx dp\, \{W_{\scriptscriptstyle
    \alpha\beta/2}(x,p)\}^2\right]\label{eq:entropies_tildeW1}
\end{eqnarray}
\end{subequations}
for an arbitrary order $\alpha$, which can easily be evaluated
explicitly with the help of Eqs.
(\ref{eq:wigner-thermal_1-3})-(\ref{eq:high-temp-regime_1}). The
resulting expressions of $\widetilde{\mathcal
W}_{\beta;\alpha}(x,p)$ and those of the R\'{e}nyi-Wigner entropy
$S_{\alpha}(\widetilde{\mathcal W}_{\beta})$ given in Eqs.
(\ref{eq:tilde_wigner1}) and (\ref{eq:wigner-entropy_2}) are clearly
straightforward to obtain but simply too large in size, and so we do
not provide them here.

\section{Evaluations of R\'{e}nyi-Wigner Entropies in the Phase Space}\label{sec:sec4}
We begin with numerical evaluations of the entropies
$S_{\alpha}(\widetilde{\mathcal W}_{\beta})$ given in Eq.
(\ref{eq:entropies_tildeW1}) for the one-dimensional box problem.
Then, we observe good agreement between
$S_{\alpha}(\widetilde{\mathcal W}_{\beta})$ and its counterpart
$S_{\alpha}(\hat{\rho}_{\scriptscriptstyle \beta}) = (1 -
\alpha)^{-1} \{(\ln Z_{\alpha\beta}) - \alpha (\ln Z_{\beta})\}$ (in
the high-temperature regime) (cf. Fig. {\ref{fig:fig6}}). It is also
interesting to compare Eq. (\ref{eq:purity_measure1}) (or Eq.
(\ref{eq:tilde_wigner1}) with $\alpha = 2$) and Eq.
(\ref{eq:purity_rel1-1}) by performing their actual evaluations for
the pure state $\widetilde{\mathcal W}_{\scriptscriptstyle 0;N}(x,p)
= (2\pi\hbar)\,\{W_{\scriptscriptstyle 0;N}(x,p)\}^2$ for Nbc as a
simple illustration of our formulation, where $W_{\scriptscriptstyle
0;N}(x,p) = (2\pi a p)^{-1} \sin\{2\,(a-|x|)\,p/\hbar\}$; for Eq.
(\ref{eq:purity_measure1}), $\mbox{Tr}(\hat{\rho}^2) = (2\pi\hbar)
\int_{-a}^a dx \int_{-\infty}^{\infty} dp\, \{W_{\scriptscriptstyle
0;N}(x,p)\}^2 = 1$ and thus $S_2(\widetilde{\mathcal
W}_{\scriptscriptstyle 0;N}) = 0$, as is easily verified. On the
other hand, Eq. (\ref{eq:purity_rel1-1}) will take the form, after
some steps of algebraic manipulations, of
\begin{eqnarray}\label{eq:purity_rel1-1-neu}
    &&(4\pi a)^{-4} \int_{-\infty}^{\infty} dx\,dp\,dx_1\,dp_1\,dx_2\,dp_2\,dx_3\,dp_3\;
    p^{-1} \left[\left\{\sin \frac{2}{\hbar}\,(a-|y_1|)\,p\right\} + \left\{y_1 \to y_4\right\}\right] \times\n\\
    &&(p_1)^{-1} \left[\left\{\sin \frac{2}{\hbar}\,(a-|y_1|)\,p_1\right\} + \left\{y_1 \to
    y_2\right\}\right] (p_2)^{-1} \left[\left\{\sin \frac{2}{\hbar}\,(a-|y_2|)\,p_2\right\} + \left\{y_2 \to
    y_3\right\}\right] \times\n\\
    &&(p_3)^{-1} \left[\left\{\sin \frac{2}{\hbar}\,(a-|y_3|)\,p_3\right\} + \left\{y_3 \to
    y_4\right\}\right]
\end{eqnarray}
as long as $|a - (\cdots)| \leq a$ in the argument of $\sin$, where
$y_1 := x + (x_1+x_2+x_3)/2; y_2 := x - (x_1-x_2-x_3)/2; y_3 := x -
(x_1+x_2-x_3)/2; y_4 := x - (x_1+x_2+x_3)/2$. Then we will have
$\int_{-\infty}^{\infty} dx \int_{-\infty}^{\infty} dx_1
\int_{-\infty}^{\infty} dx_2 \int_{-\infty}^{\infty} dx_3 \to
\int_{-a}^{a} dy \int_{-a}^{a} dy_1 \int_{-a}^{a} dy_2 \int_{-a}^{a}
dy_3$, which will finally lead to $\mbox{Tr}(\hat{\rho}^2) = 1$. As
is explicitly shown here, an evaluation of $S_2(\hat{\rho})$ via Eq.
(\ref{eq:tilde_wigner1}), considered one of our central results, is
much simpler than employing Eq. (\ref{eq:purity_rel1-1}). Besides,
we already notice that it would be a formidable task to find the
recurrence relation between entropies $S_l$ and $S_{l+1}$, with $l =
2,3,4,\cdots$ for a given distribution $\widetilde{\mathcal
W}_{\rho;\alpha}(x,p)$ (or $W_{\rho}(x,p)$) [cf. Eqs.
(\ref{eq:wigner-thermal_1-3}) and (\ref{eq:orbits-temp1})], if
needed for the analytic continuation of $l \to \alpha$.

For comparison, we briefly discuss other ``entropies'' as well,
without considering the Bopp shift. First, some moments of the
Wigner function $W_{n;D}(x,p)$ for Dbc can explicitly be evaluated
such that
\begin{subequations}
\begin{eqnarray}
    (2\pi\hbar) \int dx dp\, \{W_{n;D}(x,p)\}^2 &=& 1\label{eq:moment1}\\
    (2\pi\hbar)^2 \int dx dp\, \{W_{n;D}(x,p)\}^3 &=& \frac{1}{4} + \frac{1}{(n\pi)^2}
    \left\{\frac{15}{4} - \frac{20}{3} (-1)^n\right\}\label{eq:moment2}\\
    (2\pi\hbar)^3 \int dx dp\, \{W_{n;D}(x,p)\}^4 &=& \frac{2}{3} + \frac{25}{2
    (n\pi)^2}\label{eq:moment3}\\
    (2\pi\hbar)^4 \int dx dp\, \{W_{n;D}(x,p)\}^5 &=& \frac{23}{192} + \frac{1}{(n\pi)^2}
    \left\{\frac{475}{192} - \frac{4462}{135} (-1)^n\right\}\, +\n\\
    && \frac{1}{(n\pi)^4}
    \left\{\frac{11275}{256} + \frac{1091816}{10125}
    (-1)^n\right\}\label{eq:moment4}\,.
\end{eqnarray}
\end{subequations}
Because of $n$-dependence, Eqs.
(\ref{eq:moment2})-(\ref{eq:moment4}) cannot appropriately reflect
the higher moments $\mbox{Tr}(\hat{\rho}^3) =
\mbox{Tr}(\hat{\rho}^4) = \mbox{Tr}(\hat{\rho}^5) = 1$, as expected.
For $W_{n;N}(x,p)$ of the Nbc case, similar results will appear.
Therefore, it is obvious that the ``Wigner entropy'' given by
$S_{\mbox{\tiny W}}(\widetilde{\mathcal W}_{\scriptscriptstyle n;D})
= -\int dx dp\, \widetilde{\mathcal W}_{\scriptscriptstyle
n;D}(x,p)\,\ln\{(2\pi\hbar)\,\widetilde{\mathcal
W}_{\scriptscriptstyle 0;N}(x,p)\}$ will be $n$-dependent and so
cannot at all be used as an appropriate entropy for our purpose.
This confirms that the same will also apply for the resulting
``Wigner entropy'' $S_{\mbox{\tiny W}}(\widetilde{\mathcal
W}_{\beta}) = -\int dx dp\, \widetilde{\mathcal
W}_{\beta}(x,p)\,\ln\{(2\pi\hbar)\,\widetilde{\mathcal
W}_{\beta}(x,p)\}$ of the thermal state. On the other hand, the
Wigner entropy of the classical thermal distribution $W_{\beta}(x,p)
\to P_{\beta}(x,p) = (Z_{\beta,\mbox{\sc cl}})^{-1}\,e^{-\beta
p^2/2m}$ with $|x| < a$ is given by the closed form
\begin{equation}\label{eq:classical_wigner-entropy1}
    S_{\mbox{\tiny W}}(P_{\beta}) = -\int dx dp\,
    P_{\beta}(x,p)\,\ln\{(2\pi\hbar)\,P_{\beta}(x,p)\} = \ln(Z_{\beta,\mbox{\sc cl}}) + 1/2 +
    \ln(2\pi\hbar)\,.
\end{equation}
Likewise, by applying Eq. (\ref{eq:wigner-entropy_2}) with
substitution of $(2\pi\hbar)\,Z_{\beta} \to Z_{\beta,\mbox{\sc cl}}$
and $W_{\rho^{\scriptscriptstyle \alpha}}(x,p) \to
(2\pi\hbar)^{\scriptscriptstyle \alpha-1} (Z_{\beta,\mbox{\sc
cl}})^{\scriptscriptstyle \alpha}\,e^{-\alpha\beta p^2/2m}$, it is
straightforward to obtain the entropy $S_1(P_{\beta}) \to
\ln(Z_{\beta,\mbox{\sc cl}}) + 1/2 - \ln(2\pi\hbar)$. By setting
$(2\pi\hbar) \to 1$, we see that $S_{\mbox{\tiny W}}(P_{\beta})$ and
$S_1(P_{\beta})$ become identical.

\section{Conclusions}\label{sec:conclusion}
We have introduced the Wigner-like operational distributions
$\widetilde{\mathcal W}_{\rho;\alpha}(x,p)$ in the classical phase
space, all of which are non-negative and well-defined over the
entire phase space, by utilizing the properties of fractional
moments $\mbox{Tr}(\hat{\rho}^{\alpha})$ with $\alpha > 0$ of the
density operator $\hat{\rho}$. Then we have provided a framework for
exact evaluations of R\'{e}nyi-Wigner entropies for the
classical-like distributions $\widetilde{\mathcal
W}_{\rho;\alpha}(x,p)$, in particular for arbitrary non-Gaussian
states, which enables us to go beyond the study of R\'{e}nyi
entropies restricted to the Gaussian states with non-negative valued
Wigner functions $W_{\rho}(x,p)$. This result can be regarded as a
generalization of the preceding one developed in Ref. \cite{KIM18},
which has enabled to evaluate the entropies but essentially
restricted to integer values of $\alpha$ only.

Subsequently, we have rigorously evaluated the Wigner function
$W_{\rho}(x,p)$, directly leading to $\widetilde{\mathcal
W}_{\rho;\alpha}(x,p)$, of the thermal state of a single particle
confined by a one-dimensional infinite potential well with either
the Dirichlet or Neumann boundary condition (as a simple
non-Gaussian state), in order to illustrate our concept. We have
successfully applied our framework for this non-Gaussian state. Our
analysis has also been useful for a study of the quantal-classical
transition by expressing the density operator $\hat{\rho}$ (mixed
state) itself in terms of its phase-space counterpart
$W_{\rho}(x,p)$ (or $\widetilde{\mathcal W}_{\rho;\alpha}$) which
can be decomposed into its classical part and the quantum correction
[cf. Eqs. (\ref{eq:orbits-temp1})-(\ref{eq:high-temp-regime_1})].
This aspect will provide further insights for deeper semiclassical
analysis.

Our study will overall contribute to a better understanding of
non-Gaussian states and their transitions either in the
semiclassical limit ($\hbar \to 0$) or in the high-temperature limit
($\beta \to 0$). This phase-space approach will also be useful
information-theoretically and thermodynamically for deeper
discussions of the quantal-classical Second Law on the single
footing. We may expect that our analysis of the R{\'{e}nyi-Wigner
entropies for non-Gaussian states will contribute to making some
additional classification among non-Gaussian states and its
quantification (beyond the non-Gaussianity as their deviations from
the respective reference Gaussian states) to be pursued and also
that our approach will apply for other billiard systems (i.e.,
confined systems with different boundary shapes in two dimensions),
which are well-known to possess quantum signatures of classically
regular and chaotic motions.

\section*{Acknowledgments}
The author gratefully acknowledges the financial support provided by
the US Army Research Office (Grant No. W911NF-15-1-0145).

\appendix*\section{Derivation of Eq. (\ref{eq:wigner-thermal_1-3})}\label{sec:appendix}
%
We begin by rewriting Eqs. (\ref{eq:Wigner_thermal-1}) and
(\ref{eq:Wigner_thermal-N}) as
\begin{equation}\label{eq:Wigner_thermal-1-1}
     W_{\beta}(x,p) = \frac{1}{(2\pi\hbar)\,Z_{\beta}}
     \int_0^{y_{\scriptscriptstyle x}} dy\, \cos\left(\frac{2 a p y}{\hbar}\right)
     \sum_{n=-\infty}^{\infty} \exp(-\lambda n^2) \left\{\cos(n \pi y)
     + B \cos\left(n \pi y_x\right)\right\}\,,
\end{equation}
where $y = \xi/a$. Employing the Poisson summation rule \cite{GAS99}
\begin{equation}\label{eq:identity_2}
    \sum_{n=-\infty}^{\infty} \exp(-\lambda n^2)\, \cos(n\pi \kappa) =
    \left(\frac{\pi}{\lambda}\right)^{1/2}
    \sum_{\nu=-\infty}^{\infty} \exp\left\{-\frac{\pi^2\,(\kappa + 2\nu)^2}{4\lambda}\right\}
\end{equation}
and then with the help of the identity \cite{GRA07}
\begin{equation}\label{eq:identity_1}
    \int dn\, \exp\left(-a n^2 - 2 b n\right) = \frac{1}{2}
    \left(\frac{\pi}{a}\right)^{1/2}
    \exp\left(\frac{b^2}{a}\right)\, \mbox{erf}\left(a^{1/2}\,n +
    a^{-1/2}\,b\right)\,,
\end{equation}
Eq. (\ref{eq:Wigner_thermal-1-1}) will be transformed into
\begin{eqnarray}\label{eq:wigner-thermal_1-2}
    W_{\beta}(x,p) &=& \frac{1}{(2\pi\hbar)\,Z_{\beta}}
    \left(\frac{\pi}{\lambda}\right)^{1/2}
    \sum_{\nu=-\infty}^{\infty} \left[\int_0^{y_{\scriptscriptstyle x}} \cos\left(\frac{2 a p
    y}{\hbar}\right)\, \exp\left\{-\frac{\pi^2}{4\lambda}\left(y + 2\nu\right)^2\right\}\, dy\; +\right.\n\\
    && \left.\frac{B \hbar}{2 a p}\, \sin\left(\frac{2 a p y_x}{\hbar}\right)\,
    \exp\left\{-\frac{\pi^2}{4\lambda}\left(y_x + 2\nu\right)^2\right\}\right]\,.
\end{eqnarray}
We perform the integration over $y$ by applying the identity
(\ref{eq:identity_1}) with $n \to y$, which will result in Eq.
(\ref{eq:wigner-thermal_1-3}). In doing so, we also used, with the
help of Poisson's sum rule, the relation
\begin{equation}\label{eq:normalizing_1}
    (2\pi\hbar)\,Z_{\beta} = 2\pi\hbar \sum_{n=1}^{\infty} \exp\left(-\lambda n^2\right) = B \pi\hbar + Z_{\beta,\mbox{\sc cl}}
    \sum_{\nu=-\infty}^{\infty} \exp\left(\frac{-8 m a^2 \nu^2}{\beta \hbar^2}\right)\,,
\end{equation}
which reduces to $Z_{\beta,\mbox{\sc cl}}$ in the limit of $\hbar
\to 0$.
%

%
\newpage
\begin{figure}[htb]
\centering\hspace*{-0cm}\vspace*{-1cm}{
\includegraphics[scale=0.7]{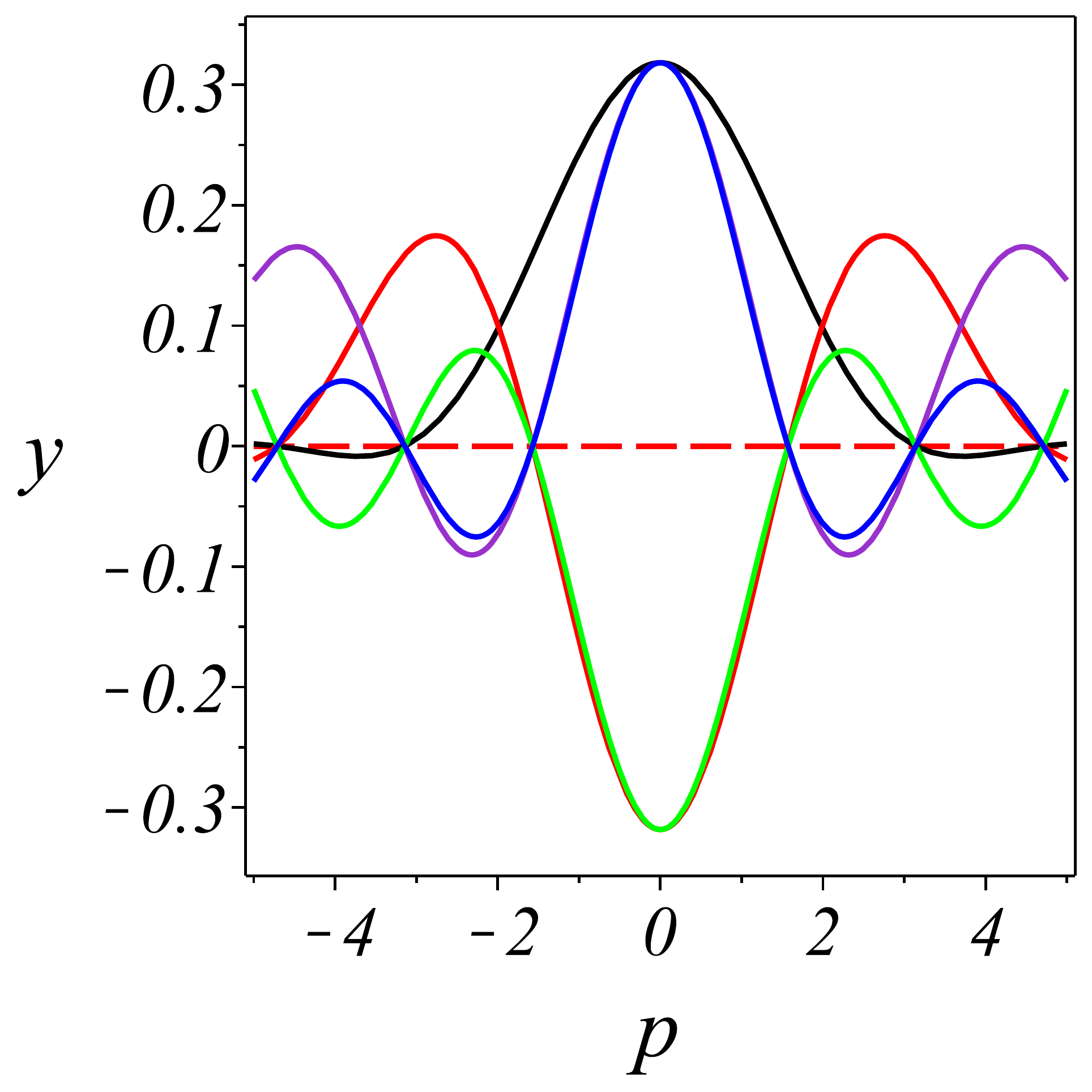}
\caption{\label{fig:fig1}}}
\end{figure}
Fig.~\ref{fig:fig1}: (Color online) Wigner functions $y =
W_{n;D}(x,p)$ for Dbc versus (dimensionless) momentum $p \equiv a
p/\hbar$ at (dimensionless) position $x \equiv x/a = 0$ with $n = 1,
2, 3, 4, 5$, each of which demonstrates negative values as its
non-Gaussian feature [cf. Eqs. (\ref{eq:wigner_box_1})]. The values
$n = 3,4,1,2,5$ are in sequence from top to bottom at $p = 5$.
\newpage
\begin{figure}[htb]
\centering\hspace*{-0cm}\vspace*{-1cm}{
\includegraphics[scale=0.7]{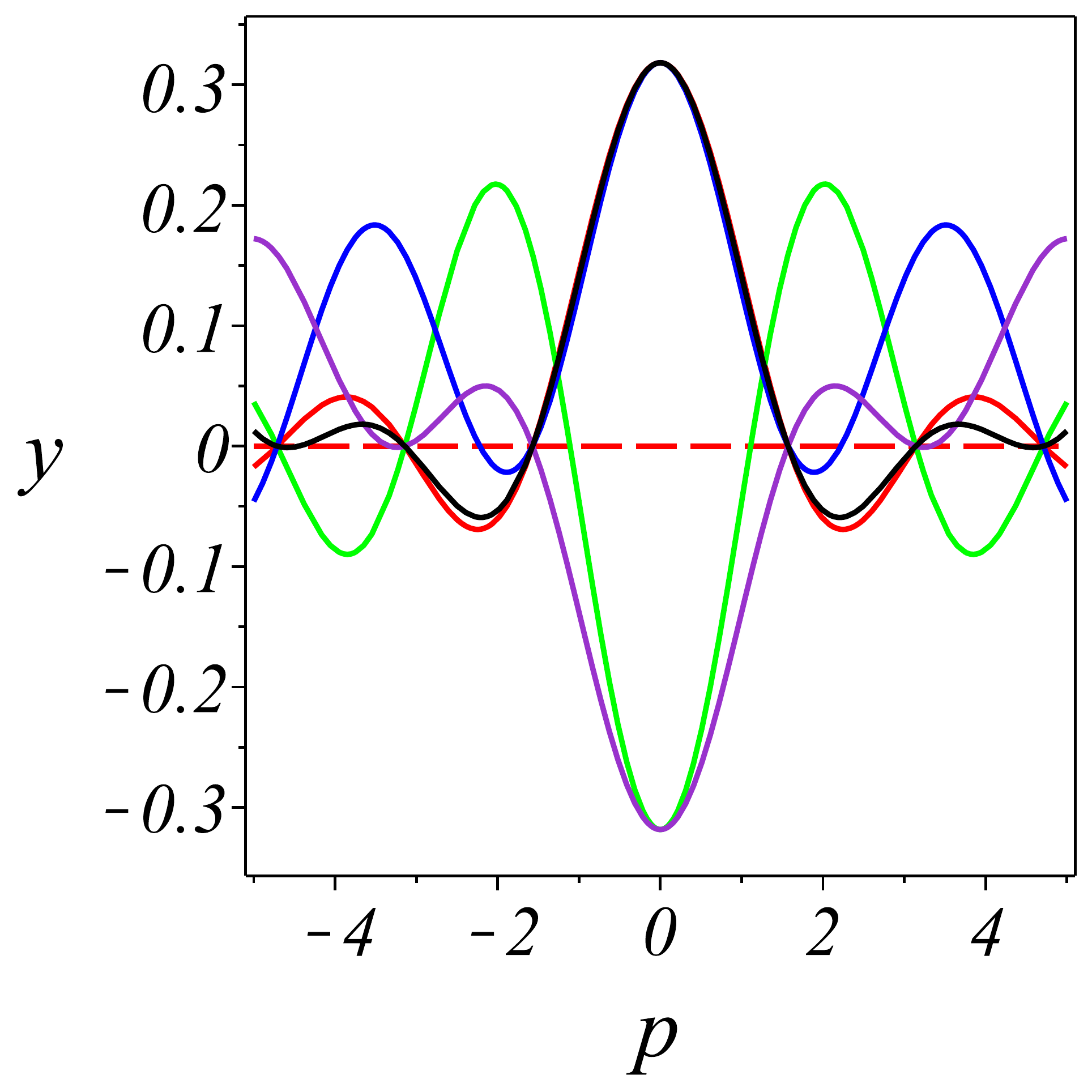}
\caption{\label{fig:fig2}}}
\end{figure}
Fig.~\ref{fig:fig2}: (Color online) Wigner functions $y =
W_{n;N}(x,p)$ for Nbc versus momentum $p$ at $x = 0$ with $n = 0, 1,
2, 3, 4$, each of which demonstrates negative values as its
non-Gaussian feature [cf. Eqs. (\ref{eq:wigner_box_n})]. The values
$n = 3,1,4,0,2$ are in sequence from top to bottom at $p = 5$.
\newpage
\begin{figure}[htb]
\centering\hspace*{-0cm}\vspace*{-0cm}{
\includegraphics[scale=0.7]{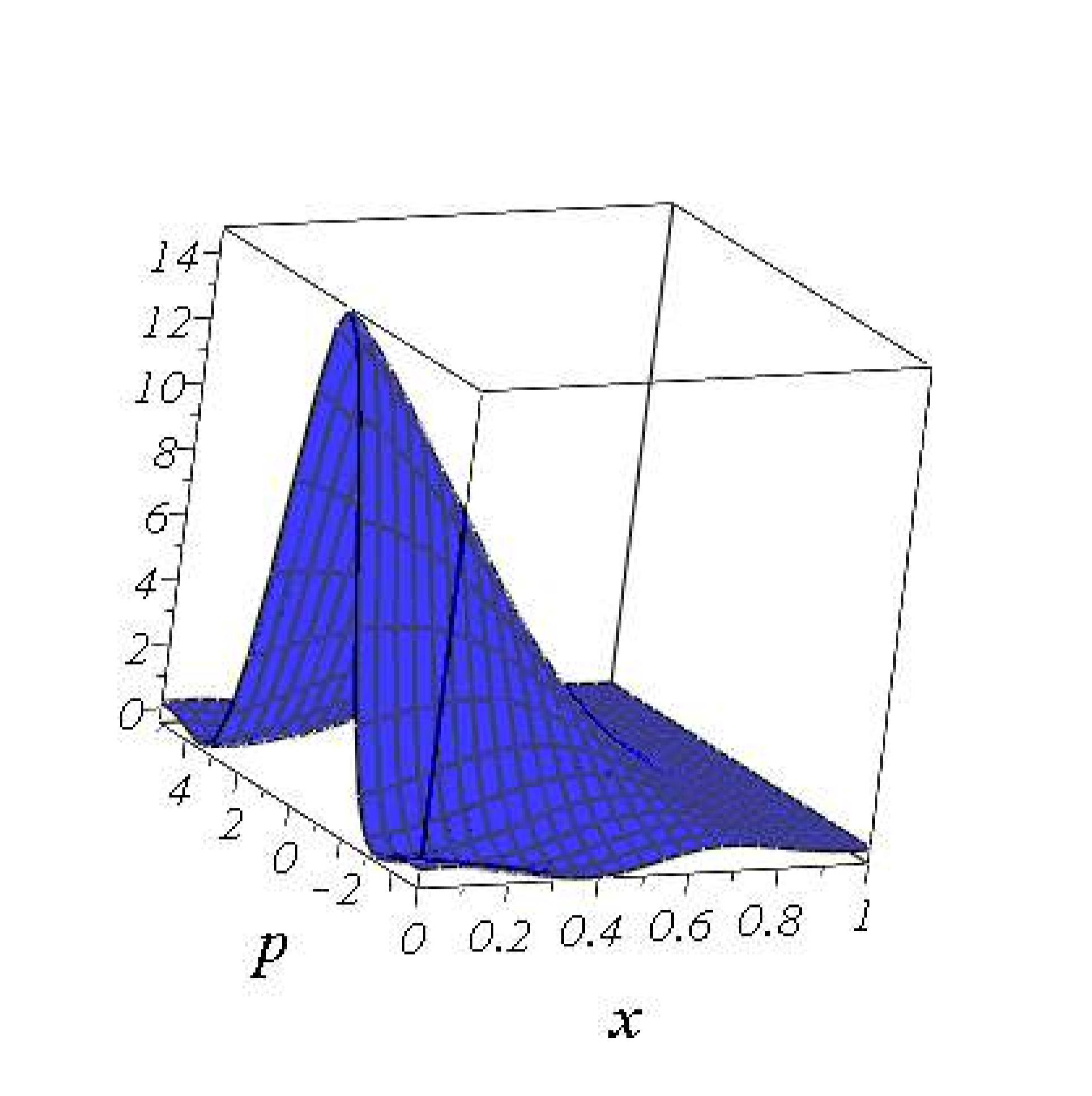}
\caption{\label{fig:fig3}}}
\end{figure}
Fig.~\ref{fig:fig3}: (Color online) The semiclassical Wigner
function $W_{\beta}(x.p)$ for Dbc versus (dimensionless) position $x
\equiv x/a$ and (dimensionless) momentum $p \equiv a p/\hbar$ [cf.
Eq. (\ref{eq:semiclassical_wigner-thermal_1-1-1})], re-scaled to
$20\, W_{\beta}(x,p)$, with the (dimensionless) temperature $\beta
\equiv \beta \hbar^2/(m a^2) = 1$. We see that $W_{\beta}(a,p) = 0$
(also $W(-a,p) = 0$ due to the symmetry of this system). Its
negative values emerge from the oscillatory behaviors around $x = 0$
(cf. Fig. \ref{fig:fig4}). Here, we have used $50$ orbits by using
the sum over $k$ with $0 \leq k \leq 50$, with good numerical
convergence. A similar result will occur for Nbc.
\newpage
\begin{figure}[htb]
\centering\hspace*{-0cm}\vspace*{-0cm}{
\includegraphics[scale=0.7]{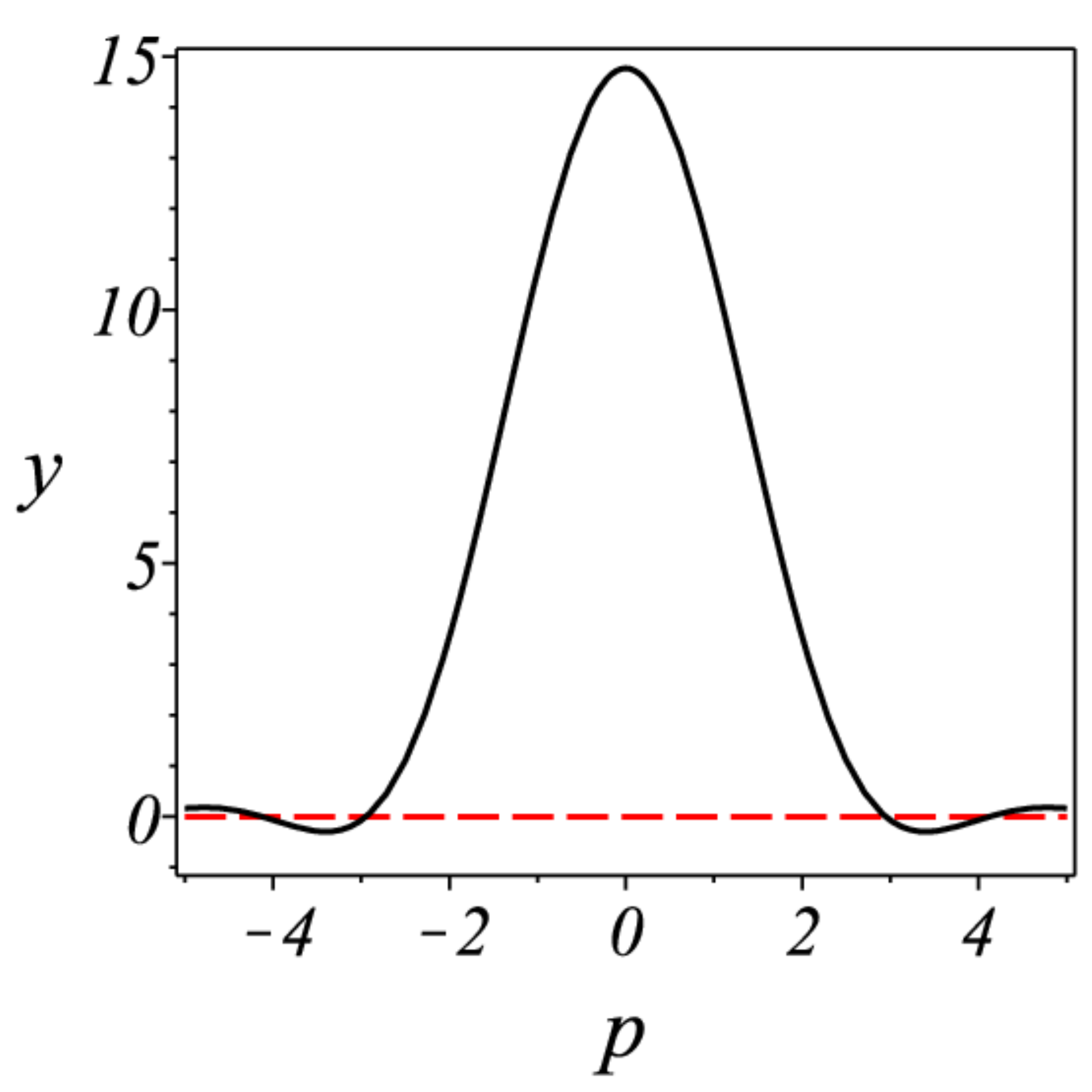}
\caption{\label{fig:fig4}}}
\end{figure}
Fig.~\ref{fig:fig4}: (Color online) The semiclassical Wigner
function $y = W_{\beta}(0,p)$ for Dbc [cf. Eq.
(\ref{eq:semiclassical_wigner-thermal_1-1-1})], re-scaled to $20\,
W_{\beta}(0,p)$, which explicitly shows its negative values, e.g.,
$W_{\beta}(0,3) = -0.0710$. Otherwise, the same parameters exist as
for Fig. \ref{fig:fig3}.
\newpage
\begin{figure}[htb]
\centering\hspace*{-3cm}\vspace*{-0cm}{
\includegraphics[scale=0.7]{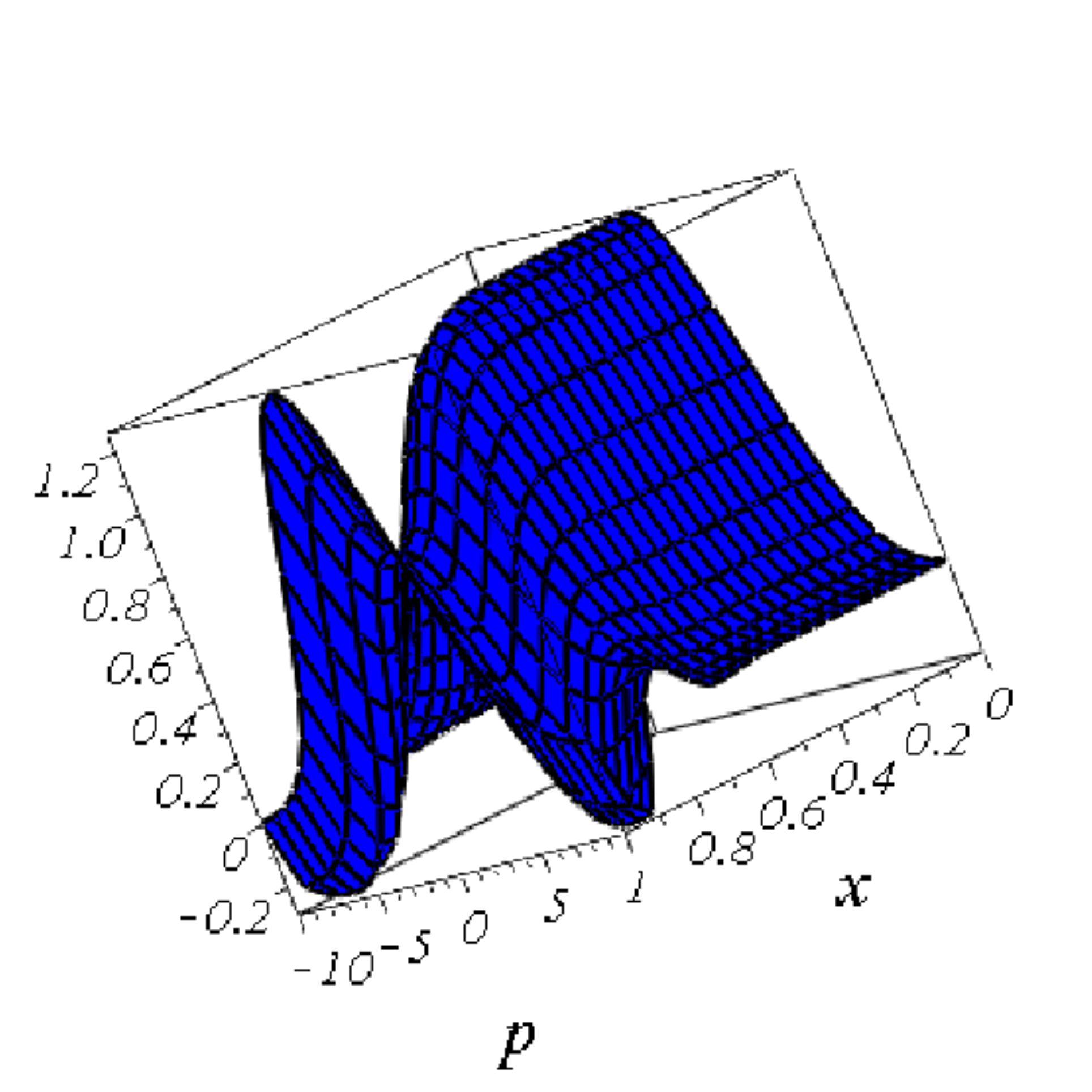}
\caption{\label{fig:fig5}}}
\end{figure}
Fig. \ref{fig:fig5}: (Color online) The Wigner function
$W_{\beta}(x.p)$ for Dbc versus (dimensionless) position $x \equiv
x/a$ and (dimensionless) momentum $p \equiv a p/\hbar$ in the
high-temperature limit [cf. Eq. (\ref{eq:high-temp-regime_1})],
re-scaled to $20\, W_{\beta}(x,p)$, with the (dimensionless)
temperature $\beta \equiv \beta \hbar^2/(m a^2) = 0.1$, which is
very high. We see that its negative values still emerge from the
highly oscillatory behaviors in the vicinity of $x = \pm a$. This
comes primarily from the contribution of the zero-length trivial
orbit corresponding to the first exponential function with $\mu =1$
in Eq. (\ref{eq:quantum-f1}). We also observe that the boundary
condition $W_{\beta}(\pm a,p) = 0$ does not hold any longer. Here,
we have used $50$ orbits by using the sum over $\mu$ with $0 \leq
\mu \leq 50$, with good numerical convergence. A similar result will
occur for Nbc.
\newpage
\begin{figure}[htb]
\centering\hspace*{-3cm}\vspace*{-0cm}{
\includegraphics[scale=0.7]{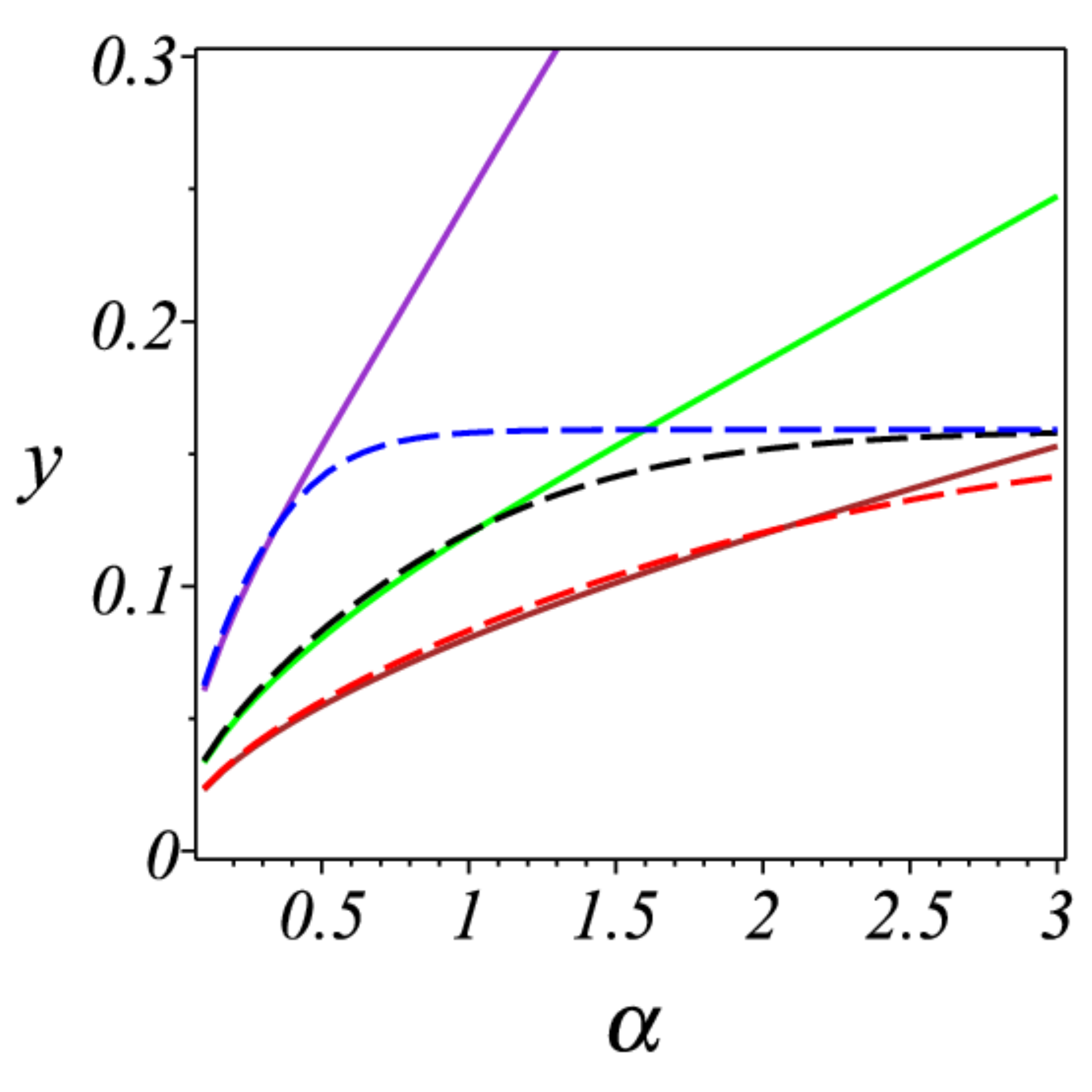}
\caption{\label{fig:fig6}}}
\end{figure}
Fig.~\ref{fig:fig6}: (Color online) Comparison between $y_1 = \int
dx dp\, \{W_{\scriptscriptstyle \alpha\beta/2}(x,p)\}^2$, exactly
computable using Eq. (\ref{eq:high-temp-regime_1}), (solid curves)
and its exact value $y_2 = (2\pi\hbar)^{-1}
(Z_{\alpha\beta})/(Z_{\scriptscriptstyle \alpha\beta/2})^2$ (dash
curves) for given orders $\alpha$ for Dbc in the high-temperature
limit, which is equivalent to the comparison between $y_1 =
S_{\alpha}(\widetilde{\mathcal W}_{\beta;\alpha})$ and its
counterpart $y_2 = S_{\alpha}(\hat{\rho}_{\scriptscriptstyle
\beta})$ in the same limit (before Eq.
(\ref{eq:purity_rel1-1-neu})); 1) the solid curves with $\beta = 3,
1, 0.5$ are in sequence from top to bottom; 2) the dash curves in
the same way. For sufficiently small values of $\alpha\beta$, we
have good agreement between the two curves for each $\beta$. Here,
we have used $15$ orbits by using the sum over $\mu$ with $0 \leq
\mu \leq 15$ in Eq. (\ref{eq:quantum-f1}), already with good
numerical convergence. A similar result will occur for Nbc. This
analysis will be useful for a study of the high-temperature
approximation in the quantum-classical transition.

\begin{thebibliography}{1}
%
\bibitem{REN61} A. R\'{e}nyi, {\em On measures of entropy and information},
in Proc. Fourth Berkeley Symp. on Math. Statist. and Prob., Vol. 1,
edited by J. Neyman (Univ. of Calif. Press, Berkeley, 1961),
547-561.
%
\bibitem{REN65} A. R\'{e}nyi, {\em On the foundations of information theory}, Rev. Intl. Stat. Inst. {\bf 33}, 1 (1965).
%
\bibitem{BEN17} A. Ben-Naim, {\em Entropy, Shannon's Measure of Information and
Boltzmann's H-Theorem}, Entropy {\bf 19}, 48 (2017).
%
\bibitem{WEH78} A. Wehrl, {\em General properties of entropy}, Rev. Mod. Phys. {\bf 50}, 221
(1978).
%
\bibitem{LIN13} N. Linden, M. Mosonyi and A. Winter, {\em The structure of R\'{e}nyi entropic
inequalities}, Proc. R. Soc. A {\bf 469}, 20120737 (2013).
%
\bibitem{MIS15} A. Misra, U. Singh, M. N. Bera and A. K. Rajagopal,
{\em Quantum Renyi relative entropies affirm universality of
thermodynamics}, Phys. Rev. E {\bf 92}, 042161 (2015).
%
\bibitem{HAM16} A. Hamma, S. M. Giampaolo and F. Illuminati, {\em Mutual information and
spontaneous symmetry breaking}, Phys. Rev. A {\bf 93}, 012303
(2016).
%
\bibitem{ABE16} S. Abe, {\em Time evolution of Renyi entropy under the Lindblad
equation}, Phys. Rev. E {\bf 94}, 022106 (2016).
%
\bibitem{KIM18} I. Kim,
{\em R\'{e}nyi $\alpha$ entropies of quantum states in closed form:
Gaussian states and a class of non-Gaussian states}, Phys. Rev. E
{\bf 97}, 062141 (2018),
%
\bibitem{DON19} X. Dong, {\em Holographic R\'{e}nyi Entropy at High Energy Density}, Phys. Rev. Lett. {\bf 122}, 041602 (2019).
%
\bibitem{GRI05} D. J. Griffiths, {\em Introduction to Quantum Mechanics} (Pearson, Upper
Saddle River, 2005).
%
\bibitem{CHU00} M. A. Nielsen and I. L. Chuang, {\em Quantum Computation and Quantum
Information} (CUP, Cambridge, 2000).
%
\bibitem{MAH04} J. Gemmer, M. Michel, and G. Mahler, {\em Quantum
Thermodynamics} (Springer, Berlin, 2004).
%
\bibitem{WIG32} E. P. Wigner, {\em On the quantum correction for
thermodynamic equilibrium}, Phys. Rev. {\bf 40}, 749 (1932).
%
\bibitem{HIL84} M. Hillery, R. F. O'Connell, M. O. Scully, and E. P. Wigner, {\em
Distribution Functions in Physics: Fundamentals}, Phys. Rep. {\bf
106}, 121 (1984).
%
\bibitem{LEE95} H.-W. Lee, {\em Theory and Application of the
Quantum Phase-Space Distribution Functions}, Phys. Rep. {\bf 259},
147 (1995).
%
\bibitem{SCH01} W. P. Schleich, {\em Quantum Optics in Phase Space}
(Wiley-VCH, Berlin, 2001).
%
\bibitem{CUR05} C. K. Zachos, D. B. Fairlie and T. L. Curtright,
{\em Quantum Mechanics in Phase Space} (World Scientific, Singapore,
2005).
%
\bibitem{BRA05} S. L. Braunstein and P. v. Loock, {\em Quantum information with continuous variables},
Rev. Mod. Phys. {\bf 77}, 513 (2005).
%
\bibitem{FER05} A. Ferraro, S. Olivares and M. G. A. Paris, {\em Gaussian States in
Quantum Information} (Bibliopolis, Napoli, 2005).
%
\bibitem{WOL06} M. M. Wolf, G. Giedke and J. I. Cirac,
{\em Extremality of Gaussian Quantum States}, Phys. Rev. Lett. {\bf
96}, 080502 (2006).
%
\bibitem{WED12} C. Weedbrook, S. Pirandola,
R. Garc\'{i}a-Patr\'{o}n, N. J. Cerf, T. C. Ralph, J. H. Shapiro and
S. Lloyd, {\em Gaussian quantum information}, Rev. Mod. Phys. {\bf
84}, 621 (2012).
%
\bibitem{OLI12} S. Olivares, {\em Quantum optics in the phase space - A tutorial on Gaussian states},
Eur. Phys. J. ST {\bf 203}, 3 (2012).
%
\bibitem{ADE14} G. Adesso, S. Ragy, and A. R. Lee, {\em Continuous Variable Quantum Information:
Gaussian States and Beyond}, Open Syst. Inf. Dyn. {\bf 21}, 1440001
(2014).
%
\bibitem{ADE12} G. Adesso, D. Girolami, and A. Serafini, {\em Measuring Gaussian Quantum Information and Correlations Using the Renyi Entropy of Order 2},
Phys. Rev. Lett. {\bf 109}, 190502 (2012).
%
\bibitem{PAT17} J. P. Santos, G. T. Landi, and M. Paternostro, {\em Wigner Entropy Production
Rate}, Phys Rev. Lett. {\bf 118}, 220601 (2017).
%
\bibitem{GEN10} M. G. Genoni, M. G. A. Paris, {\em Quantifying non-Gaussianity for
quantum information}, Phys. Rev. A {\bf 82}, 052341 (2010).
%
\bibitem{EIS02} J. Eisert, S. Scheel, and M. B. Plenio, {\em Distilling Gaussian States with Gaussian Operations is Impossible}, Phys. Rev. Lett. {\bf 89}, 137903
(2002).
%
\bibitem{GIE02} G. Giedke and J. I. Cirac, {\em Characterization of Gaussian operations and distillation of Gaussian states}, Phys. Rev. A {\bf 66}, 032316 (2002).
%
\bibitem{FIU02} J. Fiurasek, {\em Gaussian Transformations and Distillation of Entangled Gaussian States}, Phys. Rev. Lett. {\bf 89}, 137904 (2002).
%
\bibitem{RAL03} T. C. Ralph, A. Gilchrist, G. J. Milburn, W. J. Munro, and
S. Glancy, {\em Quantum computation with optical coherent states},
Phys. Rev. A {\bf 68}, 042319 (2003).
%
\bibitem{LUN08} A. P. Lund, T. C. Ralph, and H. L. Haselgrove, {\em Fault-Tolerant Linear Optical Quantum
Computing with Small-Amplitude Coherent States}, Phys. Rev. Lett.
{\bf 100}, 030503 (2008).
%
\bibitem{DEL07} F. Dell'Anno, S. D. Siena, L. Albano, and F. Illuminati, {\em Continuous-variable quantum teleportation with non-Gaussian
resources}, Phys. Rev. A {\bf 76}, 022301 (2007).
%
\bibitem{WAN15} S. Wang, L.-L. Hou, X.-F. Chen, and X.-F. Xu, {\em Continuous-variable quantum teleportation with non-Gaussian
entangled states generated via multiple-photon subtraction and
addition}, Phys. Rev. A {\bf 91}, 063832 (2015).
%
\bibitem{GEN07} M. G. Genoni, M. G. A. Paris, and K. Banaszek, {\em Measure of the
non-Gaussian character of a quantum state}, Phys. Rev. A {\bf 76},
042327 (2007).
%
\bibitem{GEN08} M. G. Genoni, M. G. A. Paris, and K. Banaszek, {\em Quantifying the
non-Gaussian character of a quantum state by quantum relative
entropy}, Phys. Rev. A {\bf 78}, 060303(R) (2008).
%
\bibitem{MAR13} P. Marian and T. A. Marian, {\em Relative entropy is an exact measure
of non-Gaussianity}, Phys. Rev. A {\bf 88}, 012322 (2013).
%
\bibitem{IVA12} J. S. Ivan, M. S. Kumar, and R. Simon, {\em A measure of non-Gaussianity for quantum states}, Quantum Inf.
Process. {\bf 11}, 853 (2012).
%
\bibitem{GHI13} I. Ghiu, P. Marian and T. A. Marian, {\em Measures of non-Gaussianity
for one-mode field states}, Phys. Scr. {\bf T153}, 014028 (2013).
%
\bibitem{BRA15} F. G. S. L. Brandao, M. Horodecki, N. H. Y. Ng, J. Oppenheim,
and S. Wehner, {\em The second laws of quantum thermodynamics},
Proc. Natl. Acad. Sci. U.S.A. {\bf 112}, 3275 (2015).
%
\bibitem{ING02} G.-L. Ingold, {\em Path Integrals and Their Application to
Dissipative Quantum Systems}, in {\em Coherent Evolution in Noisy
Environments}, edited by A. Buchleitner and K. Hornberger (Springer,
Berlin, 2002).
%
\bibitem{GNU01} S. Gnutzmann and K. \.{Z}yczkowski, {\em R\'{e}nyi-Wehrl entropies as measures of localization in
phase space}, J. Phys. A: Math. Gen. {\bf 34}, 10123 (2001).
%
\bibitem{LEE83} H.-W. Lee and M. O. Scully, {\em The Wigner Phase-Space
Description of Collision Processes}, Found. Phys. {\bf 13}, 61
(1983).
%
\bibitem{ALM90} M. A. M. de Aguiar and A. M. O. de Almeida, {\em On
the probability density interpretation of smoothed Wigner
functions}, J. Phys. A: Math. Gen. {\bf 23}, L1025 (1990).
%
\bibitem{GRO94} C. Grosche, {\em Boundary Conditions in Path
Integrals}, in Proc. Workshop {\em Singular Schr\"{o}dinger
Operators}, SISSA, Trieste (1994), and DESY Report, DESY 95-032
(1995).
%
\bibitem{DIA02} N. C. Dias and J. N. Prata, {\em Wigner functions
with boundaries}, J. Math. Phys. {\bf 43}, 4602 (2002).
%
\bibitem{FAC15} S. D. Martino and P. Facchi, {\em Quantum systems with time-dependent boundaries}, Int. J. Geom. Methods Mod. Phys.
{\bf 12}, 1560003 (2015).
%
\bibitem{GAS99} C. Gasquet and P. Witomski, {\em Fourier Analysis and Applications}
(Springer, New York, 1999).
%
\bibitem{GRA07} I. S. Gradshteyn and I. M. Ryzhik, {\em Table of Integrals, Series,
and Products}, 7th ed. (Academic Press, San Diego, 2007).
%
\bibitem{KEA87} J. P. Keating and M. V. Berry, {\em False
singularities in partial sums over closed orbits}, J. Phys. A {\bf
20}, L1139 (1987).
%
\bibitem{BRA97} M. Brack and R. K. Bhaduri, {\em Semiclassical
Physics} (Addison-Wesley, New York, 1997).
%
\bibitem{STE98} C. Grosche and F. Steiner, {\em Handbook of Feynman Path
Integrals} (Springer, Berline, 1998).
%
\bibitem{ABR65} M. Abramowitz and I. A. Stegun, {\em Handbook of Mathematical
Functions} (Dover, New York, 1965).
%
\end{thebibliography}
\end{document}